\documentclass[%
twocolumn,
superscriptaddress,
 amsmath,amssymb,
pra,
]{revtex4-2}

\usepackage[english]{babel}
\usepackage[utf8]{inputenc}
\usepackage{dcolumn}% Align table columns on decimal point
\usepackage{bm}% bold math
\usepackage{amsthm}
\usepackage{amssymb}
\usepackage{mathtools}
\usepackage{physics}
\usepackage{xcolor}
\usepackage{graphicx}
\usepackage{adjustbox}
\usepackage{placeins}
\usepackage[T1]{fontenc}
\usepackage{lipsum}
\usepackage{csquotes}
\usepackage{comment}
\usepackage{xr}
\usepackage{hyperref}

\usepackage{makecell}

\def\dEu{\ensuremath{^2\!E_u}}
\def\dEg{\ensuremath{^2\!E_g}}

\def\tA2{^{3}\!A_2}

\def\qA2{^{4}\!A_2}

\def\NV0{\rm{NV}^0}

\newcommand{\smallcaption}[1]{\caption{#1}}

\makeatletter
\newcommand*{\addFileDependency}[1]{% argument=file name and extension
  \typeout{(#1)}
  \@addtofilelist{#1}
  \IfFileExists{#1}{}{\typeout{No file #1.}}
}
\makeatother

\begin{document}

\title{Optical properties of  SiV and GeV color centers in nanodiamonds under hydrostatic pressures up to 180 GPa}

\author{Baptiste Vindolet}
\thanks{Present address: Wainvam-e, 1 Rue Galil\'ee, 56270 Ploemeur, France}
\affiliation{Universit\'e Paris-Saclay, CNRS, ENS Paris-Saclay, CentraleSupelec, LuMIn, F-91190 Gif-sur-Yvette, France}
 
\author{Marie-Pierre Adam}
\affiliation{Universit\'e Paris-Saclay, CNRS, ENS Paris-Saclay, CentraleSupelec, LuMIn, F-91190 Gif-sur-Yvette, France}

\author{Loïc Toraille}
\affiliation{Universit\'e Paris-Saclay, CNRS, ENS Paris-Saclay, CentraleSupelec, LuMIn, F-91190 Gif-sur-Yvette, France}
\affiliation{CEA DAM DIF, F-91297 Arpajon, France}
\affiliation{Université Paris-Saclay, CEA, Laboratoire Matière en Conditions Extrêmes, 91680 Bruyères-le-Châtel, France}

\author{Mayeul Chipaux}
\thanks{Present address: Institute of Physics, Ecole Polytechnique Fédérale de Lausanne (EPFL), Lausanne CH-1015, Switzerland}
\affiliation{Fachrichtung Physik, Universit\"at des Saarlandes, Campus E2.6, 66123 Saarbr\"ucken, Germany}

\author{Antoine Hilberer}
\affiliation{Universit\'e Paris-Saclay, CNRS, ENS Paris-Saclay, CentraleSupelec, LuMIn, F-91190 Gif-sur-Yvette, France}

\author{G\'eraud Dupuy}
\affiliation{Universit\'e Paris-Saclay, CNRS, ENS Paris-Saclay, CentraleSupelec, LuMIn, F-91190 Gif-sur-Yvette, France}

\author{Lukas Razinkovas}
\affiliation{Center for Physical Sciences and Technology (FTMC), Vilnius LT-10257, Lithuania}

\author{Audrius Alkauskas}
\affiliation{Center for Physical Sciences and Technology (FTMC), Vilnius LT-10257, Lithuania}

\author{Gerg\H{o} Thiering}
\affiliation{Institute for Solid State Physics and Optics, Wigner Research Centre for Physics, Budapest, POB 49, H-1525, Hungary}

\author{Adam Gali}
\affiliation{Institute for Solid State Physics and Optics, Wigner Research Centre for
Physics, Budapest, POB 49, H-1525, Hungary}
\affiliation{Department of Atomic Physics, Institute of Physics, Budapest University of Technology and Economics, M\H{u}egyetem rakpart 3., H-1111, Budapest, Hungary}

\author{Mary De Feudis}
\thanks{Present address: Laboratoire de Physique des Matériaux et Surfaces, CY Cergy Paris Université, 95031 Cergy-Pontoise, France}
\affiliation{Laboratoire des Sciences des Proc\'ed\'es et des Mat\'eriaux,  CNRS, Universit\'e Sorbonne Paris-Nord, F-93340 Villetaneuse,   France}

\author{Midrel Wilfried Ngandeu Ngambou}
\affiliation{Laboratoire des Sciences des Proc\'ed\'es et des Mat\'eriaux,  CNRS, Universit\'e Sorbonne Paris-Nord, F-93340 Villetaneuse,   France}

\author{Jocelyn Achard}
\affiliation{Laboratoire des Sciences des Proc\'ed\'es et des Mat\'eriaux,  CNRS, Universit\'e Sorbonne Paris-Nord, F-93340 Villetaneuse,   France}

\author{Alexandre Tallaire}
\affiliation{Laboratoire des Sciences des Proc\'ed\'es et des Mat\'eriaux,  CNRS, Universit\'e Sorbonne Paris-Nord, F-93340 Villetaneuse,   France}

\author{Martin Schmidt}
\affiliation{Universit\'e Paris-Saclay, CNRS, ENS Paris-Saclay, CentraleSupelec, LuMIn, F-91190 Gif-sur-Yvette, France}

\author{Christoph Becher}
\affiliation{Fachrichtung Physik, Universit\"at des Saarlandes, Campus E2.6, 66123 Saarbr\"ucken, Germany}

\author{Jean-François Roch}
\email{To whom correspondence should be addressed; E-mail: 
jean-francois.roch@ens-paris-saclay.fr}
\affiliation{Universit\'e Paris-Saclay, CNRS, ENS Paris-Saclay, CentraleSupelec, LuMIn, F-91190 Gif-sur-Yvette, France}

%\author[1]{Baptiste Vindolet}
%\email{baptiste.vindolet@ens-paris-saclay.fr}
%\affiliation{Universit\'e Paris-Saclay, CNRS, ENS Paris-Saclay, CentraleSupelec, LuMIn, F-91190 Gif-sur-Yvette, France}

\begin{abstract}

We investigate the optical properties of silicon-vacancy (SiV) and germanium-vacancy (GeV) color centers in nanodiamonds under  hydrostatic pressure up to 180 GPa. The nanodiamonds were synthetized by Si or Ge-doped plasma assisted chemical vapor deposition and, for our experiment,  pressurized in a diamond anvil cell.
Under hydrostatic pressure we observe  blue-shifts of the SiV and GeV zero-phonon lines by 17 THz (70 meV) and 78 THz (320 meV), respectively.
These measured pressure induced shifts are  in good agreement with \textit{ab initio} calculations that take into account the  lattice compression based on the  equation of state of diamond and that are extended to the case of the tin-vacancy (SnV) center.  This work provides guidance on the use of group-IV-vacancy   centers  as  quantum sensors under extreme pressures that will exploit   their specific optical and spin properties  induced by their intrinsic inversion-symmetric structure. 
%SiV and GeV centers, and more generally group-IV-vacancy (G4V) centers in diamond, are   new potential  calibration gauges associated to a practical pressure scale above the megabar.    
\end{abstract}

\maketitle

\section{Introduction}

Quantum sensing consists in using quantum systems to perform measurement of given physical quantities \cite{Degen2017}.
  Among the various systems that have been developed, the nitrogen-vacancy (NV) center in diamond has been used to demonstrate and implement a broad variety of sensing protocols, in particular for the measurement of  magnetic and electric fields, and also for the detection of stress, temperature, mechanical vibrations, and fluctuating electromagnetic fields \cite{Rondin2014}.
  Due to the stiffness of the hosting diamond crystal, the NV center is also offering a powerful solution for probing matter at the extreme static pressures that are  routinely  achieved in a diamond anvil cell (DAC) consisting of two anvils that squeeze the sample between their flattened tips \cite{Eremetsbook}, as shown in Fig. \ref{fig:Setup}a.
  In order to push the pressure limits, the pressure amplification of the DAC is optimized, essentially by reducing the culet diameter then putting a constraint on the sample size.  For pressures above 100 GPa this leads to sample chambers of less than 50~$\mu$m diameter and to the corresponding experimental difficulties in handling and signal detection. 
    %The small diameter of the tips  means that large  pressures  can  be  achieved  through the application of a small compression  force between  the anvils. However reaching pressures above 100 GPa range leads to experimental difficulties that essentially arise because the sample chamber becomes very thin, less than 10 µm, and thus  very small samples have to be loaded. 
Using the optically detected magnetic resonance (ODMR) of the NV electronic spin that can be recorded by collecting the luminescence of the NV centers through the diamond anvil, recent works have shown the applicability of NV based optical magnetometry  to micrometer sized samples under pressure and external magnetic fields \cite{Hsieh2019,lesik2019,Yip2019}. Since the NV centers can be placed in close proximity to the sample,  the    NV based detection can be used to map  the magnetic field distribution created by the sample  magnetization, with a micrometer spatial resolution  and with   a sensitivity that remains mostly unaffected by the constraints on  the sample size.

However the use of the NV center as a high pressure magnetic sensor suffers from some  limitations, such as the implementation of the microwave excitation with the constraints associated to  the DAC \cite{lesik2019} or the detrimental influence of off-axis magnetic field that may prevent its practical use at  high magnetic field \cite{tetienne2012}.  
These features are   specific to the NV center and not to diamond in general. In particular the  silicon-vacancy (SiV) and, to a somewhat lesser extent, germanium-vacancy (GeV) and tin-vacancy (SnV) centers in their negative charge state are alternative diamond point defects that have attracted considerable attention in the past decade \cite{thiering-gali-2020}.  These centers,  commonly known as group-IV-vacancy  (G4V) centers, share a nearly identical   atomic structure.  The reasons why G4V centers have potential advantages compared to the NV center are two-fold. First, G4V centers exhibit extraordinary spectral stability due to their intrinsic inversion-symmetric structure    as shown in Fig.~\ref{fig:Setup}a~\cite{gali2013,Thiering2018}. 
%~\cite{goss1996,goss2007,dhaenens-johansson2011}. 
This property is associated to the $D_{3d}$ point group symmetry of all G4V centers. 
Second,   G4V centers offer the option of all-optical, microwave free, coherent control of their spin states \cite{Pingault2014,becker2016,slyushev2017,Becker2018,Weinzetl2019,debroux2021}, allowing for applications where the use of microwave fields, as required for most NV-based sensing methods, is detrimental or technically challenging. 

As a preliminary step to envisioning the use of G4V centers for high-pressure sensing, i.e. at pressures above the megabar,   we report here  the pressure dependence of the photoluminescence (PL) spectral properties of the SiV and GeV center at room temperature, using doped nanodiamonds in a DAC. The evolution of the PL center   wavelength  with pressure for these two G4V centers is well reproduced by {\it ab initio} density functional theory (DFT) calculation of the PL zero-phonon line (ZPL). In these calculations the pressure dependence is computed by varying the lattice parameter   according to the equation of state of diamond that can be taken as a reference for the  matrix hosting the point defect \cite{Occelli2003}.

\section{Experimental set-up}

\begin{figure}
    \centering
    \includegraphics[width = \columnwidth]{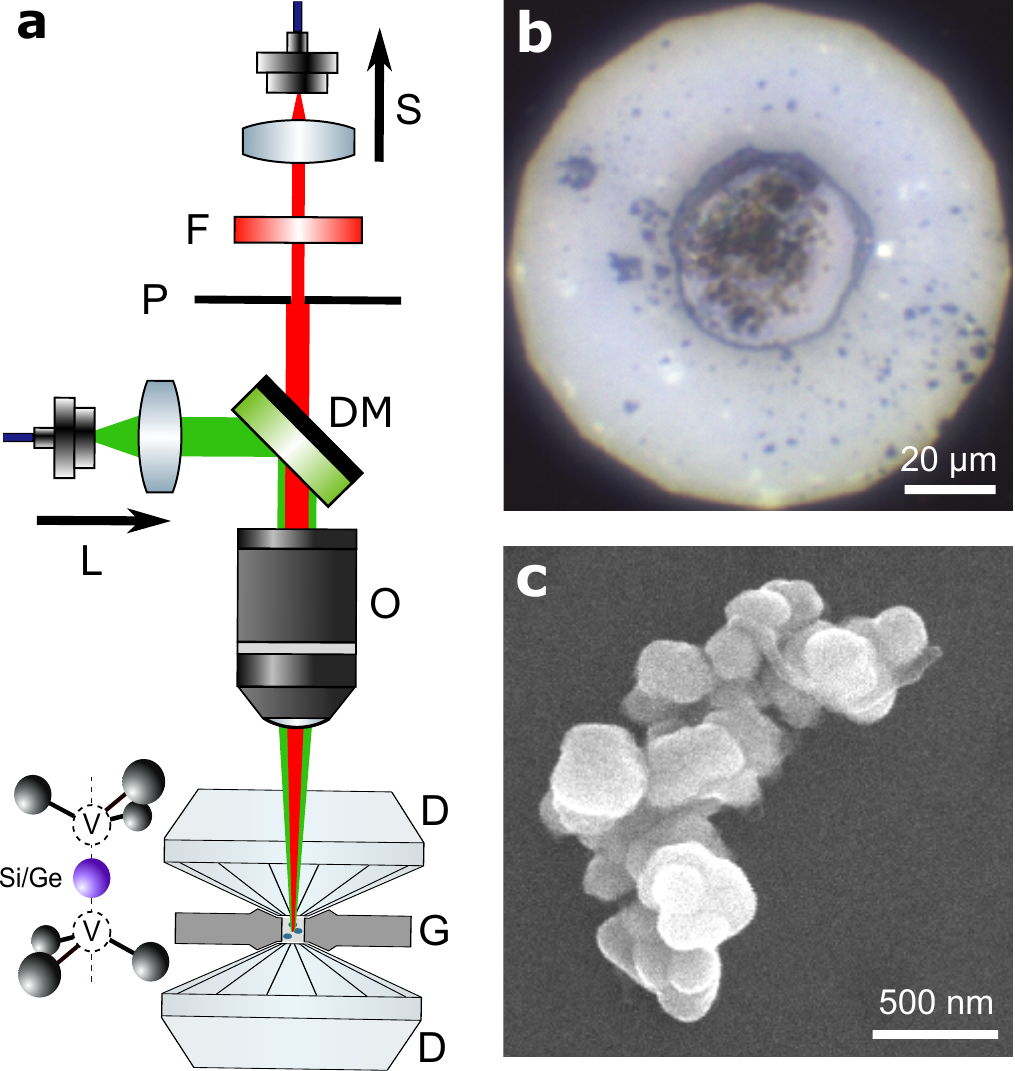}
    \smallcaption{   (\textbf{a}) Description of the high pressure experiment. SiV and GeV-doped nanodiamonds (shown in blue) are deposited on the tip of one of the two diamonds (D) assembled in a    diamond anvil cell (DAC). The geometry of the SiV and GeV center of  D$_{3d}$ symmetry is shown on the left. The Si/Ge impurity (in purple) is linked to two vacancies (in dashed lines) aligned along one crystallographic axis [111] of the  lattice of carbon atoms (in black).  
    %The DAC pressurized chamber  corresponds to the volume between the  flattened tips of the anvils  and the hole in the  metallic gasket (G) that creates the lateral confinement. 
    The excitation laser  (L) is transmitted through an optical fiber, reflected by a dichroic mirror (DM) and then focused on the anvil tip with a  microscope objective (O). 
    The SiV/GeV photoluminescence (PL) is  collected by the same objective, transmitted by the dichroic mirror and focused in an optical fiber linked to  a photon counting detector or a spectrograph (S). The remaining parasitic light  from L is removed by a longpass filter (F).   (\textbf{b}) Image of the DAC metallic gasket (G), observed through the diamond anvil, at 8.1~GPa during the GeV spectroscopy experiment.  (\textbf{c}) Scanning electron microscopy  image of a typical cluster of GeV-doped nanodiamonds.
    }    
    \label{fig:Setup}
\end{figure}

The pressure dependence of the G4V center PL spectra    is investigated using nanodiamonds  doped with   SiV and GeV centers. The nanodiamonds were  synthesized by plasma assisted chemical vapor deposition (CVD) while introducing inside the plasma a solid silicon or germanium source in the vicinity  of a molybdenum holder \cite{Feudis2020}. The amount of SiV and GeV centers directly produced in the nanodiamonds can be controlled by adding N$_2$ and O$_2$ gases to the standard mixture of H$_2$/CH$_4$ used for the CVD diamond growth \cite{tallaire2019}. This method is well suited for the mass production of nanodiamonds doped with SiV or GeV centers. We estimate the incorporated color center density to a few tens of ppb,  corresponding to several hundreds of color centers per nanodiamond with a typical size of about 200 nm.   The nanodiamonds are then retrieved by rinsing the holder with   ethanol. A drop of $\lesssim 1~\mu$L of each of these  solutions is then deposited on the flattened  tips of  diamond anvils.  The anvils, made of synthetic  ultrapure diamond, are cut with {\it Almax-Boehler} design  \cite{boehleralmax2004}. The anvil tip is a (100) cristallographic plane  and has a diameter  of 100 $\mu$m. After the evaporation of the ethanol solvent,  clusters of the CVD-grown nanodiamonds  appear on the tip, as shown in Figs.~\ref{fig:Setup}b and \ref{fig:Setup}c.

  The anvils are  mounted in a non-magnetic DAC with a rhenium metallic gasket ensuring  lateral confinement  (Fig. \ref{fig:Setup}a).
  The DAC is  loaded with neon gas as the pressure transmitting medium. The soft neon environment, which becomes solid   at $4.7\, {\rm GPa}$ \cite{neonhighpressure1991},  ensures the hydrostatic compression of the nanodiamonds.  The pressure in the DAC can be continuously tuned using a metallic membrane that controls the load applied on the anvils~\cite{letoullec1988}.
  The DAC is then integrated in a customized optical confocal microscope  equipped with a microscope objective (0.35 numerical  aperture and 18 mm working distance) that collects the PL of the G4V centers through the diamond anvil on which the nanodiamonds have been deposited.

\section{Spectroscopy of the  silicon-vacancy center}

PL spectra of seven clusters of SiV-doped nanodiamonds were recorded for  increasing pressures up to 180 GPa where the failure of an anvil induced an irreversible  decompression of the DAC. The PL was excited using a   single-mode cw laser at 532 nm wavelength. A  ruby crystal of micrometer size was introduced in the DAC  as a pressure gauge~\cite{shen2020}. The  pressure in the DAC was  simultaneously determined from the shifts of the ruby PL and of the first-order Raman mode of the diamond anvil under load, following the analysis described in Ref. \cite{akahama2004}. 

\begin{figure}
    %\centering
    \includegraphics[width=\linewidth]{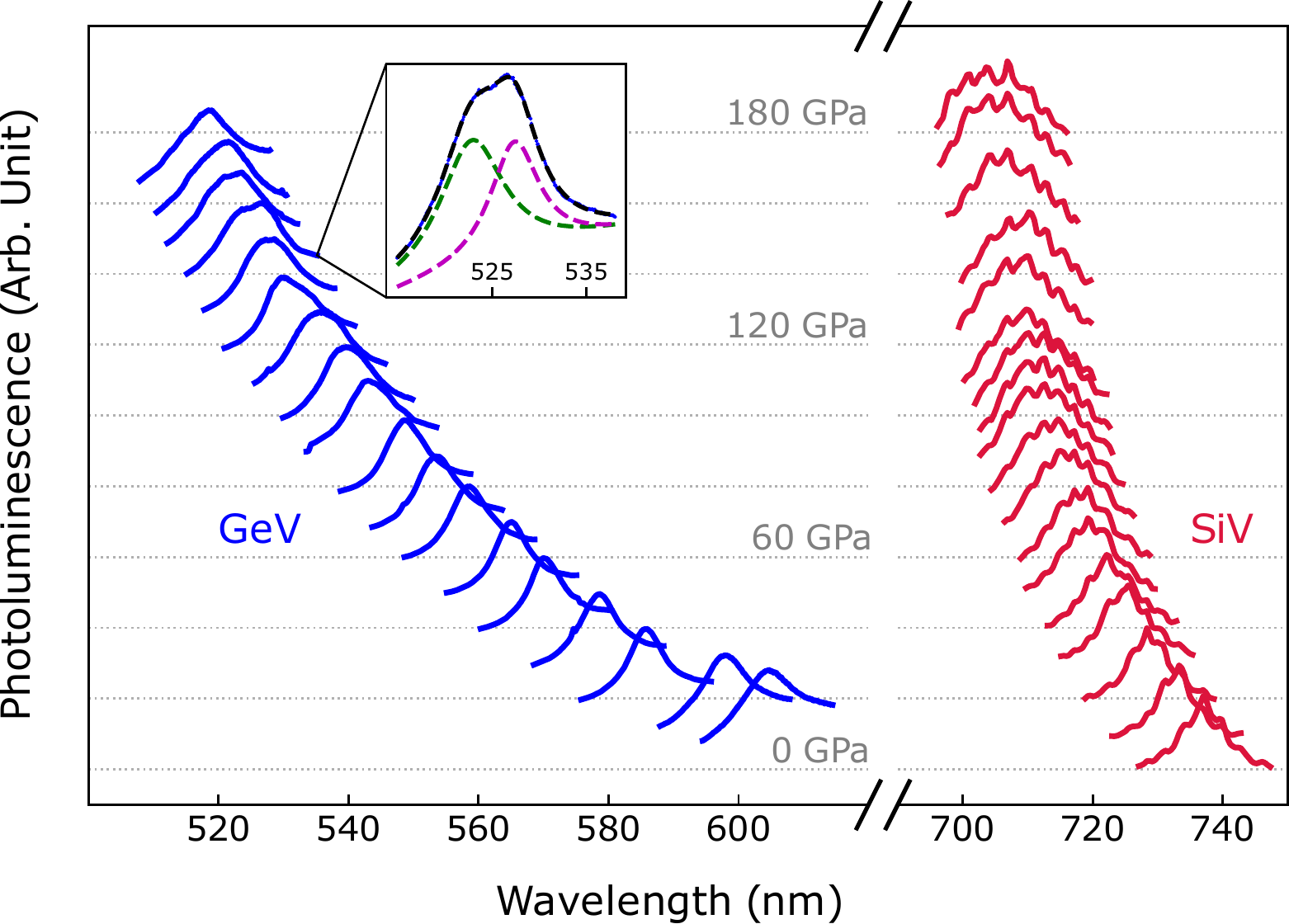}
    \smallcaption{Pressure dependence  of the  photoluminescence (PL) spectra of SiV centers (shown in red, right) and GeV centers (shown in blue, left) in CVD grown nanodiamonds.  The modulation on the SiV spectra is an artifact due to parasitic interference on the imaging detector that we first used to record the PL spectrum of the nanodiamonds. The GeV spectra were recorded with an improved spectrograph free of this artifact.  The increase of pressure   induces a blue shift and a broadening of the PL lines.  Pressure steps for SiV (from lowest to highest curve): 1, 9, 20, 32, 40, 51, 60, 70, 79, 89, 96, 103, 110, 125, 137, 155, 171, and 180 GPa. Pressure steps for GeV (from lowest to highest curve): 8, 12, 20, 30, 40, 50, 60, 69, 79, 90, 99, 109, 119, 130, 140, 149, 157, and 168 GPa.  
    The plots are separated vertically by the change of pressure which is indicated by the horizontal, dotted lines.  Inset: GeV spectrum recorded at 140 GPa  showing its deconvolution by two shifted  Lorentzian components.}
    \label{fig:ZPLwithPressure}
\end{figure}

Fig.~\ref{fig:ZPLwithPressure} shows the spectra recorded on one cluster of SiV-doped nanodiamonds for increasing pressure steps of about 10~GPa.  The increase of pressure induces a blue shift of the PL line as  expected from the enhanced confinement of the SiV electronic wavefunctions in the ground and excited states.  The spectra are then fitted by a Lorentzian function in order to determine both the center emission wavelength and the linewidth. 
 Recorded with a constant laser excitation power of 15 mW, the total PL intensity associated to a given nanodiamond slightly decreased   with increasing pressure. We attribute this effect to the shift of the absorption spectrum, leading to a decrease of the absorption cross-section at 532 nm wavelength.

The pressure dependence  of the center  energy, averaged over the seven aggregates that were investigated, is shown on Fig.~\ref{fig:ZPL_exp_theory}. We estimated the error  of the pressure measurement as $\pm 1$ GPa   below 79 GPa, where the pressure was accurately determined using the Ruby reference,  and $\pm 8$ GPa   for the Raman scattering pressure measurement that was used   above 79 GPa. The statistical error of the SiV mean energy is about $\pm 2.5$ meV for 95\% confidence interval.  The graph also indicates the corresponding value of  the lattice parameter inferred from the equation of state of diamond  under hydrostatic condition \cite{Occelli2003}. %then showing an almost linear dependence.
The results are in good agreement with previous measurements where the   pressure dependence of the SiV PL was measured up to 50~GPa using HPHT grown  nanodiamonds loaded in a DAC~\cite{lyapin2018}. 
  Although the size of the anvil tip prevented us to record a significant number of points at the start of the pressure exploration, the shift as a function of pressure below 20 GPa is approximately linear with a slope of  about 1 meV/GPa. This value  agrees with independent measurements previously performed in this weak strain regime by  bending  a diamond cantilever that integrated a single SiV
center~\cite{meesala2018}.  

Fig.~\ref{fig:ZPLwithPressure} also shows that the pressure induced shift is associated to  a broadening of the PL.  Indeed the {\it ab initio} DFT calculation on the strain induced  ZPL shift done in Ref.~\cite{lindner2018} showed that hydrostatic pressure results in a blue shift whereas uniaxial stress results in red shifts.
The broad inhomogeneous distribution of the  spectral properties of SiV centers that is observed in milled or CVD grown nanodiamonds was then  explained by the  uncontrolled strain environment compared to the highly reproducible properties of  SiV centers designed by ion implantation in high quality bulk diamond crystals. Such inhomogeneities could be reduced by the synthesis of high-pressure high-temperature nanodiamonds  with Si or Ge precursors  \cite{davydov2014,chen2018} or by improving the CVD growth process \cite{Isa2022}. The PL linewidth also depends on the electron-phonon interaction. Since the electronic states are shifted and mixed by strain, the electron-phonon scattering might be affected (see e.g. Ref.~\cite{lindner2018}) then possibly contributing to a pressure-dependent homogeneous broadening of the PL spectrum.

\section{Spectroscopy of the  germanium-vacancy center}

A similar  and complementary experiment was 
performed by loading a DAC with   GeV-doped nanodiamonds   and  recording the evolution of their PL with increasing pressures up to  about 170~GPa.  The DAC was prepared following the same procedure as   previously   without integrating a ruby crystal since its PL could induce a parasitic signal superimposed with the GeV PL. The pressure in the DAC was then determined from the first-order Raman mode associated to the compression of the diamond lattice, using the same analysis as in the SiV experiment  \cite{akahama2004}.

 %The optical components (dichroic mirror and high-pass filter) of the confocal microscope were  adapted accordingly to the modification of the excitation wavelength. 
 %At 160~GPa,  a parasitic signal centered around 550~nm wavelength suddenly appeared. This effect could be associated to the onset of  cracks which could anticipate the break of the anvil that indeed happened when the pressure reached 168 GPa.  

Fig.~\ref{fig:ZPLwithPressure} shows the spectra recorded on a given cluster of GeV-doped nanodiamonds, exhibiting  a similar behavior as for the SiV-doped nanodiamonds. From 8 GPa to 93 GPa, the   GeV centers were  excited using a single-mode cw laser at 532~nm wavelength  and with a power of 75 mW.  Above this pressure,   the laser excitation was switched to a  single-mode cw  laser at 488~nm wavelength in order to compensate the blue shift of the GeV absorption spectrum.   The laser excitation power was 37 mW.  The PL spectrum could then be recorded until the failure of the anvils that happened  at 168 GPa.  Note that  above approximately 100 GPa, the PL line is split in  two  components as shown in the inset of figure~ \ref{fig:ZPLwithPressure}. This splitting could be induced by the non-hydrostatic stress inside the nanodiamond which  then breaks   the $D_{3d}$    symmetry that characterizes the G4V center~\cite{hughes1967,hepp2014},
 similarly to previous measurements realized on the NV center \cite{grazioso2013}.  The experimental spectra were then fitted by two Lorentzian functions and the PL center energy was taken as the average of the two center positions of this fit. 

Fig.~\ref{fig:ZPL_exp_theory} shows the evolution of the PL center  energy, averaged over the three clusters of GeV-doped nanodiamonds that were investigated, as a function of the pressure  in the DAC. The pressure was determined with an uncertainty of     $\pm 1$ GPa   and the    statistical error of the GeV mean energy is about $\pm 10$ meV for 95\% confidence interval. The  shift  with pressure is about four times faster for the GeV center than   for the SiV center.
% We expect a similar splitting on the SiV center ZPL for the same reasons, but we did not observe it experimentally, certainly because of the modulations of the SiV spectra we obtained experimentally (figure~\ref{fig:ZPLwithPressure}-a).
% \cite{kaplyanskii1959,kaplyanskii1964} ?

\section{\textit{Ab initio} study of ZPL energies as a function of pressure}

\subsection{Methodology}

The electronic structure and the optical excitation energies of the
G4V centers in diamond can be efficiently computed using the
spin-polarized DFT formalism. In the molecular-orbital picture, the
ground $\dEg$ and excited $\dEu$ states of these point defects can be
expressed as single Slater determinant wavefunctions with respective
electronic configurations of $e_u^4e_g^3$ and
$e_u^3e_g^4$~\cite{gali2013}. Therefore, in the framework of Kohn–Sham
DFT, the energy and geometric structure of the excited $\dEu$ state
can be calculated by employing the so-called
delta-self-consistent-field ($\Delta$SCF) approximation, whereby one
$e_u$ electron in the lower-lying occupied Kohn--Sham level is
promoted to an empty $e_g$ level (see, e.g.,
Refs.~\mbox{\cite{gali2013,Thiering2018,londero2018}}).

To describe the SiV and the GeV centers' electronic structure and
extend it to the case of the SnV center, we used the SCAN
exchange–correlation functional~\cite{scan} that belongs to the class
of so-called meta-GGA functionals. This functional provides an
accurate description of bulk diamond structural properties, yielding
predictions of the diamond lattice constant $a_0=3.554~\mbox{\AA}$ and bulk modulus (based on the
Rose–Vinet equation of state~\cite{eos}) $B=460~\mbox{GPa}$   in close
agreement with the experimental values
$a_0=3.555~\mbox{\AA}$~\cite{pan2012a} and
$B=446~\mbox{GPa}$~\cite{zouboulis1998}. The point defects were then modeled
using $4\times 4\times 4$ supercells with 512 atomic sites, and the
Brillouin zone was sampled at the $\Gamma$ point. We used the
projector-augmented wave (PAW) approach~\cite{paw} as implemented in
the Vienna Ab-initio Simulation Package (VASP)~\cite{vasp} with a
plane-wave energy cutoff of 600~eV. Calculations for non-zero stresses
are then performed by modifying the lattice constant of the defected
cell according to the equation of state of diamond~\cite{eos, Occelli2003} using
theoretical parameters calculated for bulk diamond.

\begin{figure}[b]
  % \centering
  \includegraphics[width=1\linewidth]{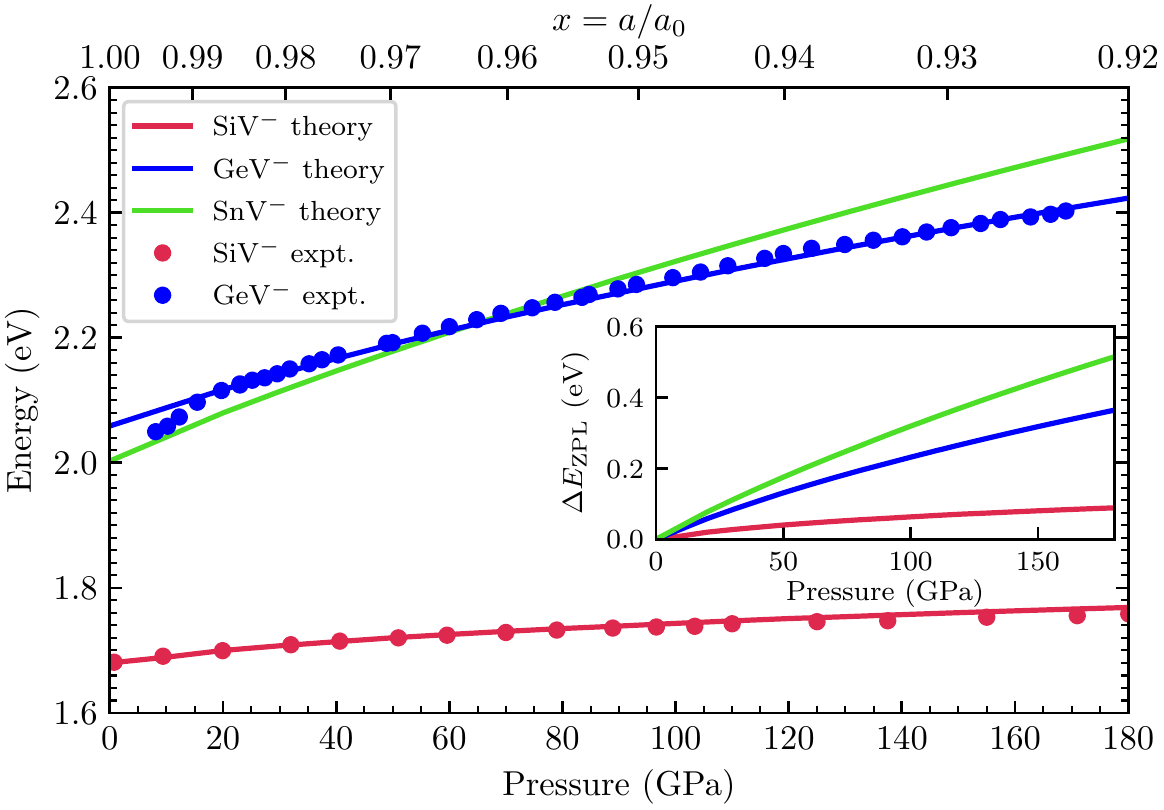}
  \caption{Values of the PL center energies obtained from the
    experimental measurements of Fig.~2 for the SiV and GeV centers,
    plotted as a function of pressure.  The error bars    indicated in the main text    are smaller than the size of the data points. The   upper axis indicates the corresponding values of the
    lattice parameter ratio $x=a/a_0$, where $a$ is the value  deduced from the equation of state
    of diamond at a given pressure~\cite{Occelli2003} and $a_0$ is the value at null pressure~\cite{eos}.
    The data
    for the SiV and GeV centers are represented by blue (lower) and
    red (upper) dots, respectively. 
   The two solid lines show the
    results of the \textit{ab initio} calculations of the zero-phonon
    line (ZPL) energy, extended to the case of the SnV center (in green). The
    calculated lines have been shifted to retrieve the ZPL energies at
    null pressure (1.68~eV for the SiV center, 2.06~eV for the GeV
    center, and 2.00~eV for the SnV center, respectively). The deviation observed for the GeV center below 20 GPa is attributed to    a bias in the pressure estimate   when we started to increase the pressure load applied to the DAC. The measurement of pressure using the Raman scattering signal is relevant  only above 20 GPa, where we observe a good agreement with the result of the ab-initio calculations.     The inset
    shows the relative shift of the ZPL energies as calculated from the total energy differences between excited and ground states, with the
    influence of the growing atomic radii of the group-IV atom once
    embedded in the diamond lattice as an interstitial
    impurity.
    \label{fig:ZPL_exp_theory}}
\end{figure}

\subsection{Results}

The ZPL energies at each pressure are calculated as the difference between the total energies of the excited and ground states using the $\Delta$SCF approximation. At zero pressure, we obtain ZPL energies of 1.57 eV, 2.00 eV, and 1.98 eV for the negatively charged SiV, GeV, and SnV centers, respectively. These values are slightly smaller than the experimentally measured ZPL energies of 1.68~eV, 2.06~eV, and 2.00~eV. Note that the accuracy of the $\Delta$SCF method with SCAN functional is close to that of the HSE functional \cite{Thiering2018}, albeit at much lower computational costs.

The calculated ZPL values of G4V centers as a pressure function are shown in   Fig.~\ref{fig:ZPL_exp_theory} by solid lines. For a meaningful
comparison with an experiment, a constant offset is applied to the theoretical curves  so that the ZPL energies at null pressure correspond to the
reported experimental values. When aligned this way, the DFT results
agree very well with the center energies that were previously measured
for the SiV and the GeV centers in all the pressure range that was  probed in the experiments.

\begin{figure}
  % \centering
  \includegraphics[width=1\linewidth]{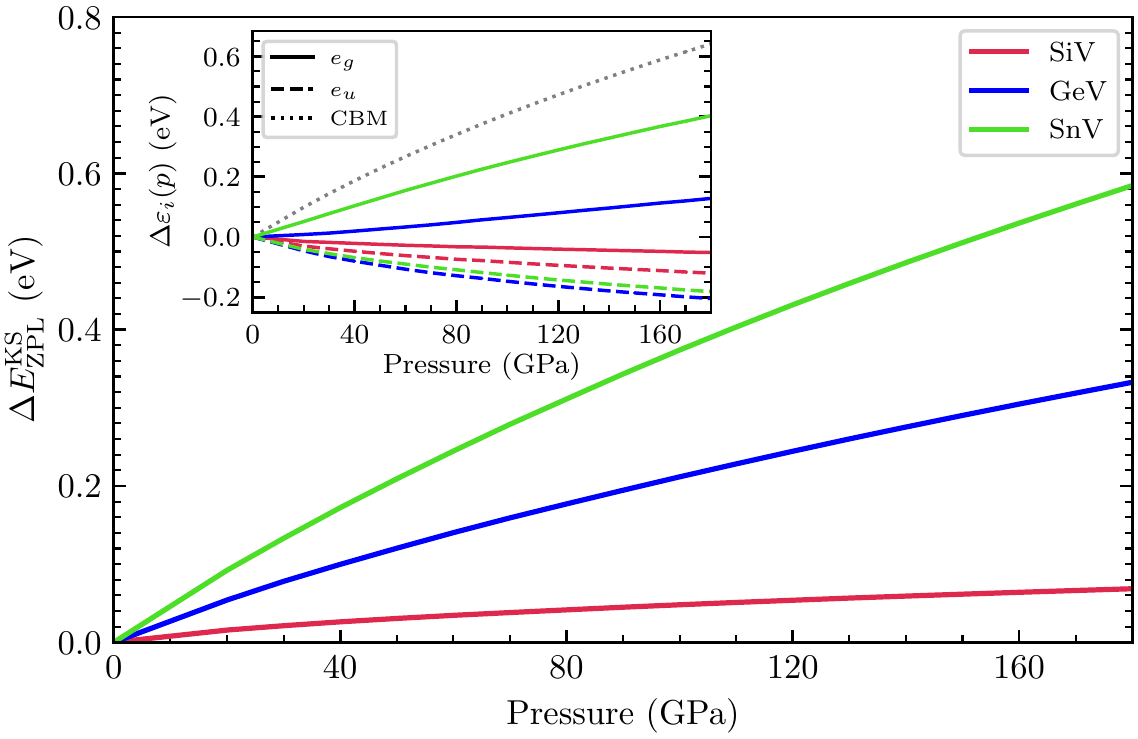}
  \caption{Relative shift of the ZPL energies as a function of
    pressure as calculated from Kohn--Sham eigenvalues
    $\Delta E_{\mathrm{ZPL}}^{\mathrm{KS}}(p) = \varepsilon_{e_g}(p) -
    \varepsilon_{e_u}(p)$. The inset shows the relative change of the
    single-particle Kohn--Sham levels in the reference system of
    valence band maximum (VBM). The $e_g$ and $e_u$ single-particle
    energies are represented as solid and dashed lines. Dotted gray
    line denotes the conduction band minimum (CBM) as calculated in
    the $\Gamma$ point of the $4\times 4\times4$ supercell.}
  \label{fig:ks_zpl}
\end{figure}

The inset of Fig.~\ref{fig:ZPL_exp_theory}  shows the relative change of
theoretical ZPL energy ($\Delta$ZPL) for the three different G4V centers as a
function of pressure. We see that the rate of change is more
pronounced for heavier impurity atoms with higher atomic radii. Keeping in mind the similarity
of the electronic structure of all G4V defects, this naturally prompts
a question regarding the reason for this difference.

\subsection{Single-particle picture}

Inspection of Kohn--Sham molecular orbital states reveals that the
trend can be explained qualitatively using a single-particle
picture. Indeed the single-particle energies allow for a qualitative chemical insight since the excitation energy can be roughly approximated as a difference between unoccupied and occupied states. 

Fig.~\ref{fig:ks_zpl} shows the relative change of the
ZPL as a function of pressure as calculated from a difference between
$e_g$ and $e_u$ single-particle energies in the electronic ground
state. To investigate the orbital energy response to strain, the
inset of Fig.~\ref{fig:ks_zpl} shows the change of single-particle
energies in the reference system of valence band maximum (VBM)
$\Delta \varepsilon_i(p) \equiv [\varepsilon_i(p) -
\varepsilon_{\mathrm{VBM}}(p)] - [\varepsilon_i(0) -
\varepsilon_{\mathrm{VBM}}(0)]$. Here, $\varepsilon_i(p)$ is the
Kohn--Sham energy of orbital $\varepsilon_i$ at pressure $p$, and
$\varepsilon_{\mathrm{VBM}}$ is the orbital energy of VBM.\@ One can
observe that the major difference between all three G4V centers lies
in the deformation potential of the $e_g$ orbital. Based on the
argument that the VBM and conduction band minimum (CBM) states are of bonding and antibonding
characters~\cite{harrison2012}, one can deduce that $e_g$ orbital changes
its nature from bonding to antibonding as we increase the atomic
number of the group-IV atom. An extended analysis of the difference in
the electronic structure and chemical bonding of G4V centers will be
given in an upcoming study.

\section{Summary and conclusion}

We measured the pressure dependence of the SiV and GeV PL   to about 180~GPa by   integrating  nanodiamonds in a DAC  that were   doped with Si or Ge impurities during their plasma-assisted growth. The experimental results  are in good agreement with  \textit{ab initio} calculations that   compute the pressure dependence of the ZPL of these G4V centers to the equation of state of diamond. This direct link suggests that  SiV and GeV centers  could be used as      pressure calibration gauges, adding a new tool to the implementation of a practical pressure scale above the megabar \cite{shen2020}. 
%This optical pressure measurement can be facilitated by the fact that   nanodiamonds can  host a high concentration of G4V centers, leading to  bright and spectrally narrow photoluminescence that can be efficiently excited and collected through one anvil of the DAC.    

The pressure dependence of the PL energies also reveals a  difference between the G4V centers,    being more rigidly bonded  in the diamond crystal lattice for increasing atomic radii. These measurements are the first steps to determine the optical and magnetic properties of G4V centers under  a broad set of parameters that could be  explored by combining high pressure and low temperature:   Jahn-Teller relaxation energies, spin-orbit splitting of the ground and excited states under the influence of high stress, orbital relaxation rate. Lastly, charge-neutral G4V centers, and in particular the
neutral SiV$^0$ center \cite{Rose2018}, could be a complementary resource since these defects can host a  coherent spin that can be optically addressed with near infrared light. \\

\section{Acknowledgments}

We are grateful to Paul Loubeyre for his insightful advices and comments, and to Florent Occelli  for his constant help  and availability in handling    high pressure instrumentation. {\bf Funding:}  This work has received funding from the  QuantERA II program under  project no. 4991922368 SENSEXTREME (grant S-QUANTERA-22-1 from the Research Council of Lithuania),  from  the Agence Nationale de la Recherche under the project SADAHPT and the  ESR/EquipEx+ program (grant number ANR-21-ESRE-0031), from the EMPIR program co-financed by the Participating States and  the European Union’s Horizon 2020 research and innovation program (20IND05 QADeT),  and from the Paris \^Ile-de-France R\'egion in the framework of DIM SIRTEQ. 
 Computations were performed on the supercomputer GALAX of the Center for Physical Sciences and Technology, Lithuania, and on the High-Performance Computing Center ``HPC Sauletekis'' in the Faculty of Physics, Vilnius University. 
A.G.\ acknowledges the National Research, Development, and Innovation Office of Hungary (NKFIH) grant No.\ KKP129866 of the National Excellence Program of Quantum-coherent materials project and the Quantum Information National Laboratory supported by the Ministry of Innovation and Technology of Hungary.  The work of BV was funded by a Ph.D. research grant from  D\'el\'egation G\'en\'erale de l'Armement.  JFR acknowledges support from Institut Universitaire de France. {\bf Data and material availability:} All experimental data described here are available at Zenodo \cite{zenodog4v}. 

\section{Appendix}

\begin{table}[h!] 
    \centering
    \begin{tabular}{ | c | c | c | c |  }
    \hline
   & SiV center & GeV center & SnV center \\
   \hline \hline 
Measured   & \makecell{$1.0$ \\ (0-20  {\rm GPa}) } &  \makecell{$2.7$  \\ (20-40   {\rm GPa})}   &   \\
   \hline
   \makecell{Computed \\ (below $50 \, {\rm GPa}$) }  & 1.00   & 2.90   &  3.85   \\
   \hline
   \makecell{Ref. \cite{lyapin2018} \\  (below $50 \, {\rm GPa}$) }  & 1.09  &  &  \\
   \hline
   \makecell{Ref. \cite{lyapin2018:GeV} \\ (below $6 \, {\rm GPa}$)}   &    & 3.11 &  \\
   \hline
   \makecell{Ref. \cite{razgulov2021} \\ (below $9 \, {\rm GPa}$)} &     &  & 3.52 \\
    \hline
    \makecell{Ref. \cite{ekimov2019} \\ (DFT)} & 
   1.1   & 3.2 & 3.5 \\
    \hline
    \end{tabular}
    \caption{Pressure dependence of the measured PL mean energies and computed ZPL energies for the G4V centers. The linear dependence is given   in meV/GPa. Taking into account the bias in the pressure measurement at the start of the pressure load, the experimental value  for the GeV center is determined  between 20 to 40 GPa.
    Extrapolation  below 20 GPa might give a result closer to the computed value due to   the nonlinearity of the diamond equation of state.   }
    \label{tab:shift}
\end{table}

Table \ref{tab:shift} compares our measured and  computed values of the linear dependence of the PL mean energies and ZPL energies in the low pressure regime (up to 20 GPa)   with previously published values. The experimental values were obtained by integrating in a DAC powders of Si \cite{lyapin2018}, Ge \cite{lyapin2018:GeV}  and Sn \cite{razgulov2021} doped microcrystals that were synthetized from hydrocarbons at high pressures and high temperatures.  The table also indicates the pressure shift that was previously computed  using  the Quantum ESPRESSO package implemented with a computational periodic supercell consisting of 83 atomic sites \cite{ekimov2019}.

\bibliographystyle{apsrev4-2}
\bibliography{Bibliography.bib}

%apsrev4-2.bst 2019-01-14 (MD) hand-edited version of apsrev4-1.bst
%Control: key (0)
%Control: author (72) initials jnrlst
%Control: editor formatted (1) identically to author
%Control: production of article title (-1) disabled
%Control: page (0) single
%Control: year (1) truncated
%Control: production of eprint (0) enabled
\begin{thebibliography}{46}%
\makeatletter
\providecommand \@ifxundefined [1]{%
 \@ifx{#1\undefined}
}%
\providecommand \@ifnum [1]{%
 \ifnum #1\expandafter \@firstoftwo
 \else \expandafter \@secondoftwo
 \fi
}%
\providecommand \@ifx [1]{%
 \ifx #1\expandafter \@firstoftwo
 \else \expandafter \@secondoftwo
 \fi
}%
\providecommand \natexlab [1]{#1}%
\providecommand \enquote  [1]{``#1''}%
\providecommand \bibnamefont  [1]{#1}%
\providecommand \bibfnamefont [1]{#1}%
\providecommand \citenamefont [1]{#1}%
\providecommand \href@noop [0]{\@secondoftwo}%
\providecommand \href [0]{\begingroup \@sanitize@url \@href}%
\providecommand \@href[1]{\@@startlink{#1}\@@href}%
\providecommand \@@href[1]{\endgroup#1\@@endlink}%
\providecommand \@sanitize@url [0]{\catcode `\\12\catcode `\$12\catcode
  `\&12\catcode `\#12\catcode `\^12\catcode `\_12\catcode `\%12\relax}%
\providecommand \@@startlink[1]{}%
\providecommand \@@endlink[0]{}%
\providecommand \url  [0]{\begingroup\@sanitize@url \@url }%
\providecommand \@url [1]{\endgroup\@href {#1}{\urlprefix }}%
\providecommand \urlprefix  [0]{URL }%
\providecommand \Eprint [0]{\href }%
\providecommand \doibase [0]{https://doi.org/}%
\providecommand \selectlanguage [0]{\@gobble}%
\providecommand \bibinfo  [0]{\@secondoftwo}%
\providecommand \bibfield  [0]{\@secondoftwo}%
\providecommand \translation [1]{[#1]}%
\providecommand \BibitemOpen [0]{}%
\providecommand \bibitemStop [0]{}%
\providecommand \bibitemNoStop [0]{.\EOS\space}%
\providecommand \EOS [0]{\spacefactor3000\relax}%
\providecommand \BibitemShut  [1]{\csname bibitem#1\endcsname}%
\let\auto@bib@innerbib\@empty
%</preamble>
\bibitem [{\citenamefont {Degen}\ \emph {et~al.}(2017)\citenamefont {Degen},
  \citenamefont {Reinhard},\ and\ \citenamefont {Cappellaro}}]{Degen2017}%
  \BibitemOpen
  \bibfield  {author} {\bibinfo {author} {\bibfnamefont {C.~L.}\ \bibnamefont
  {Degen}}, \bibinfo {author} {\bibfnamefont {F.}~\bibnamefont {Reinhard}},\
  and\ \bibinfo {author} {\bibfnamefont {P.}~\bibnamefont {Cappellaro}},\
  }\href {https://doi.org/10.1103/revmodphys.89.035002} {\bibfield  {journal}
  {\bibinfo  {journal} {Reviews of Modern Physics}\ }\textbf {\bibinfo {volume}
  {89}},\ \bibinfo {pages} {035002} (\bibinfo {year} {2017})},\ \Eprint
  {https://arxiv.org/abs/1611.02427} {1611.02427} \BibitemShut {NoStop}%
\bibitem [{\citenamefont {Rondin}\ \emph {et~al.}(2014)\citenamefont {Rondin},
  \citenamefont {Tetienne}, \citenamefont {Hingant}, \citenamefont {Roch},
  \citenamefont {Maletinsky},\ and\ \citenamefont {Jacques}}]{Rondin2014}%
  \BibitemOpen
  \bibfield  {author} {\bibinfo {author} {\bibfnamefont {L.}~\bibnamefont
  {Rondin}}, \bibinfo {author} {\bibfnamefont {J.-P.}\ \bibnamefont
  {Tetienne}}, \bibinfo {author} {\bibfnamefont {T.}~\bibnamefont {Hingant}},
  \bibinfo {author} {\bibfnamefont {J.-F.}\ \bibnamefont {Roch}}, \bibinfo
  {author} {\bibfnamefont {P.}~\bibnamefont {Maletinsky}},\ and\ \bibinfo
  {author} {\bibfnamefont {V.}~\bibnamefont {Jacques}},\ }\href
  {https://doi.org/10.1088/0034-4885/77/5/056503} {\bibfield  {journal}
  {\bibinfo  {journal} {Reports on Progress in Physics}\ }\textbf {\bibinfo
  {volume} {77}},\ \bibinfo {pages} {056503} (\bibinfo {year}
  {2014})}\BibitemShut {NoStop}%
\bibitem [{\citenamefont {Eremets}(1996)}]{Eremetsbook}%
  \BibitemOpen
  \bibfield  {author} {\bibinfo {author} {\bibfnamefont {M.~I.}\ \bibnamefont
  {Eremets}},\ }\href@noop {} {\emph {\bibinfo {title} {High Pressure
  Experimental Methods}}}\ (\bibinfo  {publisher} {Oxford University Press},\
  \bibinfo {address} {Oxford},\ \bibinfo {year} {1996})\BibitemShut {NoStop}%
\bibitem [{\citenamefont {Hsieh}\ \emph {et~al.}(2019)\citenamefont {Hsieh},
  \citenamefont {Bhattacharyya}, \citenamefont {Zu}, \citenamefont {Mittiga},
  \citenamefont {Smart}, \citenamefont {Machado}, \citenamefont {Kobrin},
  \citenamefont {Höhn}, \citenamefont {Rui}, \citenamefont {Kamrani},
  \citenamefont {Chatterjee}, \citenamefont {Choi}, \citenamefont {Zaletel},
  \citenamefont {Struzhkin}, \citenamefont {Moore}, \citenamefont {Levitas},
  \citenamefont {Jeanloz},\ and\ \citenamefont {Yao}}]{Hsieh2019}%
  \BibitemOpen
  \bibfield  {author} {\bibinfo {author} {\bibfnamefont {S.}~\bibnamefont
  {Hsieh}}, \bibinfo {author} {\bibfnamefont {P.}~\bibnamefont
  {Bhattacharyya}}, \bibinfo {author} {\bibfnamefont {C.}~\bibnamefont {Zu}},
  \bibinfo {author} {\bibfnamefont {T.}~\bibnamefont {Mittiga}}, \bibinfo
  {author} {\bibfnamefont {T.~J.}\ \bibnamefont {Smart}}, \bibinfo {author}
  {\bibfnamefont {F.}~\bibnamefont {Machado}}, \bibinfo {author} {\bibfnamefont
  {B.}~\bibnamefont {Kobrin}}, \bibinfo {author} {\bibfnamefont {T.~O.}\
  \bibnamefont {Höhn}}, \bibinfo {author} {\bibfnamefont {N.~Z.}\ \bibnamefont
  {Rui}}, \bibinfo {author} {\bibfnamefont {M.}~\bibnamefont {Kamrani}},
  \bibinfo {author} {\bibfnamefont {S.}~\bibnamefont {Chatterjee}}, \bibinfo
  {author} {\bibfnamefont {S.}~\bibnamefont {Choi}}, \bibinfo {author}
  {\bibfnamefont {M.}~\bibnamefont {Zaletel}}, \bibinfo {author} {\bibfnamefont
  {V.~V.}\ \bibnamefont {Struzhkin}}, \bibinfo {author} {\bibfnamefont {J.~E.}\
  \bibnamefont {Moore}}, \bibinfo {author} {\bibfnamefont {V.~I.}\ \bibnamefont
  {Levitas}}, \bibinfo {author} {\bibfnamefont {R.}~\bibnamefont {Jeanloz}},\
  and\ \bibinfo {author} {\bibfnamefont {N.~Y.}\ \bibnamefont {Yao}},\ }\href
  {https://doi.org/10.1126/science.aaw4352} {\bibfield  {journal} {\bibinfo
  {journal} {Science}\ }\textbf {\bibinfo {volume} {366}},\ \bibinfo {pages}
  {1349} (\bibinfo {year} {2019})}\BibitemShut {NoStop}%
\bibitem [{\citenamefont {Lesik}\ \emph {et~al.}(2019)\citenamefont {Lesik},
  \citenamefont {Plisson}, \citenamefont {Toraille}, \citenamefont {Renaud},
  \citenamefont {Occelli}, \citenamefont {Schmidt}, \citenamefont {Salord},
  \citenamefont {Delobbe}, \citenamefont {Debuisschert}, \citenamefont
  {Rondin}, \citenamefont {Loubeyre},\ and\ \citenamefont {Roch}}]{lesik2019}%
  \BibitemOpen
  \bibfield  {author} {\bibinfo {author} {\bibfnamefont {M.}~\bibnamefont
  {Lesik}}, \bibinfo {author} {\bibfnamefont {T.}~\bibnamefont {Plisson}},
  \bibinfo {author} {\bibfnamefont {L.}~\bibnamefont {Toraille}}, \bibinfo
  {author} {\bibfnamefont {J.}~\bibnamefont {Renaud}}, \bibinfo {author}
  {\bibfnamefont {F.}~\bibnamefont {Occelli}}, \bibinfo {author} {\bibfnamefont
  {M.}~\bibnamefont {Schmidt}}, \bibinfo {author} {\bibfnamefont
  {O.}~\bibnamefont {Salord}}, \bibinfo {author} {\bibfnamefont
  {A.}~\bibnamefont {Delobbe}}, \bibinfo {author} {\bibfnamefont
  {T.}~\bibnamefont {Debuisschert}}, \bibinfo {author} {\bibfnamefont
  {L.}~\bibnamefont {Rondin}}, \bibinfo {author} {\bibfnamefont
  {P.}~\bibnamefont {Loubeyre}},\ and\ \bibinfo {author} {\bibfnamefont
  {J.-F.}\ \bibnamefont {Roch}},\ }\href
  {https://doi.org/10.1126/science.aaw4329} {\bibfield  {journal} {\bibinfo
  {journal} {Science}\ }\textbf {\bibinfo {volume} {366}},\ \bibinfo {pages}
  {1359} (\bibinfo {year} {2019})}\BibitemShut {NoStop}%
\bibitem [{\citenamefont {Yip}\ \emph {et~al.}(2019)\citenamefont {Yip},
  \citenamefont {Ho}, \citenamefont {Yu}, \citenamefont {Chen}, \citenamefont
  {Zhang}, \citenamefont {Kasahara}, \citenamefont {Mizukami}, \citenamefont
  {Shibauchi}, \citenamefont {Matsuda}, \citenamefont {Goh},\ and\
  \citenamefont {Yang}}]{Yip2019}%
  \BibitemOpen
  \bibfield  {author} {\bibinfo {author} {\bibfnamefont {K.~Y.}\ \bibnamefont
  {Yip}}, \bibinfo {author} {\bibfnamefont {K.~O.}\ \bibnamefont {Ho}},
  \bibinfo {author} {\bibfnamefont {K.~Y.}\ \bibnamefont {Yu}}, \bibinfo
  {author} {\bibfnamefont {Y.}~\bibnamefont {Chen}}, \bibinfo {author}
  {\bibfnamefont {W.}~\bibnamefont {Zhang}}, \bibinfo {author} {\bibfnamefont
  {S.}~\bibnamefont {Kasahara}}, \bibinfo {author} {\bibfnamefont
  {Y.}~\bibnamefont {Mizukami}}, \bibinfo {author} {\bibfnamefont
  {T.}~\bibnamefont {Shibauchi}}, \bibinfo {author} {\bibfnamefont
  {Y.}~\bibnamefont {Matsuda}}, \bibinfo {author} {\bibfnamefont {S.~K.}\
  \bibnamefont {Goh}},\ and\ \bibinfo {author} {\bibfnamefont {S.}~\bibnamefont
  {Yang}},\ }\href {https://doi.org/10.1126/science.aaw4278} {\bibfield
  {journal} {\bibinfo  {journal} {Science}\ }\textbf {\bibinfo {volume}
  {366}},\ \bibinfo {pages} {1355} (\bibinfo {year} {2019})}\BibitemShut
  {NoStop}%
\bibitem [{\citenamefont {Tetienne}\ \emph {et~al.}(2012)\citenamefont
  {Tetienne}, \citenamefont {Rondin}, \citenamefont {Spinicelli}, \citenamefont
  {Chipaux}, \citenamefont {Debuisschert}, \citenamefont {Roch},\ and\
  \citenamefont {Jacques}}]{tetienne2012}%
  \BibitemOpen
  \bibfield  {author} {\bibinfo {author} {\bibfnamefont {J.-P.}\ \bibnamefont
  {Tetienne}}, \bibinfo {author} {\bibfnamefont {L.}~\bibnamefont {Rondin}},
  \bibinfo {author} {\bibfnamefont {P.}~\bibnamefont {Spinicelli}}, \bibinfo
  {author} {\bibfnamefont {M.}~\bibnamefont {Chipaux}}, \bibinfo {author}
  {\bibfnamefont {T.}~\bibnamefont {Debuisschert}}, \bibinfo {author}
  {\bibfnamefont {J.-F.}\ \bibnamefont {Roch}},\ and\ \bibinfo {author}
  {\bibfnamefont {V.}~\bibnamefont {Jacques}},\ }\href
  {https://doi.org/10.1088/1367-2630/14/10/103033} {\bibfield  {journal}
  {\bibinfo  {journal} {New Journal of Physics}\ }\textbf {\bibinfo {volume}
  {14}},\ \bibinfo {pages} {103033} (\bibinfo {year} {2012})}\BibitemShut
  {NoStop}%
\bibitem [{\citenamefont {Thiering}\ and\ \citenamefont
  {Gali}(2020)}]{thiering-gali-2020}%
  \BibitemOpen
  \bibfield  {author} {\bibinfo {author} {\bibfnamefont {G.}~\bibnamefont
  {Thiering}}\ and\ \bibinfo {author} {\bibfnamefont {A.}~\bibnamefont
  {Gali}},\ }in\ \href@noop {} {\emph {\bibinfo {booktitle} {Diamond for
  Quantum Applications Part 1}}},\ \bibinfo {series} {Semiconductors and
  Semimetals}, Vol.\ \bibinfo {volume} {103},\ \bibinfo {editor} {edited by\
  \bibinfo {editor} {\bibfnamefont {R.}~\bibnamefont {Mendelsohn}}\ and\
  \bibinfo {editor} {\bibfnamefont {J.~E.}\ \bibnamefont {Neumann}}}\ (\bibinfo
   {publisher} {Academic Press},\ \bibinfo {year} {2020})\ Chap.~\bibinfo
  {chapter} {1}, pp.\ \bibinfo {pages} {1--36}\BibitemShut {NoStop}%
\bibitem [{\citenamefont {Gali}\ and\ \citenamefont {Maze}(2013)}]{gali2013}%
  \BibitemOpen
  \bibfield  {author} {\bibinfo {author} {\bibfnamefont {A.}~\bibnamefont
  {Gali}}\ and\ \bibinfo {author} {\bibfnamefont {J.~R.}\ \bibnamefont
  {Maze}},\ }\href {https://doi.org/10.1103/physrevb.88.235205} {\bibfield
  {journal} {\bibinfo  {journal} {Physical Review B}\ }\textbf {\bibinfo
  {volume} {88}},\ \bibinfo {pages} {235205} (\bibinfo {year} {2013})},\
  \Eprint {https://arxiv.org/abs/1310.2137} {1310.2137} \BibitemShut {NoStop}%
\bibitem [{\citenamefont {Thiering}\ and\ \citenamefont
  {Gali}(2018)}]{Thiering2018}%
  \BibitemOpen
  \bibfield  {author} {\bibinfo {author} {\bibfnamefont {G.}~\bibnamefont
  {Thiering}}\ and\ \bibinfo {author} {\bibfnamefont {A.}~\bibnamefont
  {Gali}},\ }\href {https://doi.org/10.1103/physrevx.8.021063} {\bibfield
  {journal} {\bibinfo  {journal} {Physical Review X}\ }\textbf {\bibinfo
  {volume} {8}},\ \bibinfo {pages} {021063} (\bibinfo {year} {2018})},\ \Eprint
  {https://arxiv.org/abs/1804.07004} {1804.07004} \BibitemShut {NoStop}%
\bibitem [{\citenamefont {Pingault}\ \emph {et~al.}(2014)\citenamefont
  {Pingault}, \citenamefont {Becker}, \citenamefont {Schulte}, \citenamefont
  {Arend}, \citenamefont {Hepp}, \citenamefont {Godde}, \citenamefont
  {Tartakovskii}, \citenamefont {Markham}, \citenamefont {Becher},\ and\
  \citenamefont {Atatüre}}]{Pingault2014}%
  \BibitemOpen
  \bibfield  {author} {\bibinfo {author} {\bibfnamefont {B.}~\bibnamefont
  {Pingault}}, \bibinfo {author} {\bibfnamefont {J.~N.}\ \bibnamefont
  {Becker}}, \bibinfo {author} {\bibfnamefont {C.}~\bibnamefont {Schulte}},
  \bibinfo {author} {\bibfnamefont {C.}~\bibnamefont {Arend}}, \bibinfo
  {author} {\bibfnamefont {C.}~\bibnamefont {Hepp}}, \bibinfo {author}
  {\bibfnamefont {T.}~\bibnamefont {Godde}}, \bibinfo {author} {\bibfnamefont
  {A.~I.}\ \bibnamefont {Tartakovskii}}, \bibinfo {author} {\bibfnamefont
  {M.}~\bibnamefont {Markham}}, \bibinfo {author} {\bibfnamefont
  {C.}~\bibnamefont {Becher}},\ and\ \bibinfo {author} {\bibfnamefont
  {M.}~\bibnamefont {Atatüre}},\ }\href
  {https://doi.org/10.1103/PhysRevLett.113.263601} {\bibfield  {journal}
  {\bibinfo  {journal} {Physical Review Letters}\ }\textbf {\bibinfo {volume}
  {113}},\ \bibinfo {pages} {263601} (\bibinfo {year} {2014})}\BibitemShut
  {NoStop}%
\bibitem [{\citenamefont {Becker}\ \emph {et~al.}(2016)\citenamefont {Becker},
  \citenamefont {Görlitz}, \citenamefont {Arend}, \citenamefont {Markham},\
  and\ \citenamefont {Becher}}]{becker2016}%
  \BibitemOpen
  \bibfield  {author} {\bibinfo {author} {\bibfnamefont {J.~N.}\ \bibnamefont
  {Becker}}, \bibinfo {author} {\bibfnamefont {J.}~\bibnamefont {Görlitz}},
  \bibinfo {author} {\bibfnamefont {C.}~\bibnamefont {Arend}}, \bibinfo
  {author} {\bibfnamefont {M.}~\bibnamefont {Markham}},\ and\ \bibinfo {author}
  {\bibfnamefont {C.}~\bibnamefont {Becher}},\ }\href
  {https://doi.org/10.1038/ncomms13512} {\bibfield  {journal} {\bibinfo
  {journal} {Nature Communications}\ }\textbf {\bibinfo {volume} {7}},\
  \bibinfo {pages} {13512} (\bibinfo {year} {2016})},\ \Eprint
  {https://arxiv.org/abs/1603.00789} {1603.00789} \BibitemShut {NoStop}%
\bibitem [{\citenamefont {Siyushev}\ \emph {et~al.}(2017)\citenamefont
  {Siyushev}, \citenamefont {Metsch}, \citenamefont {Ijaz}, \citenamefont
  {Binder}, \citenamefont {Bhaskar}, \citenamefont {Sukachev}, \citenamefont
  {Sipahigil}, \citenamefont {Evans}, \citenamefont {Nguyen}, \citenamefont
  {Lukin}, \citenamefont {Hemmer}, \citenamefont {Palyanov}, \citenamefont
  {Kupriyanov}, \citenamefont {Borzdov}, \citenamefont {Rogers},\ and\
  \citenamefont {Jelezko}}]{slyushev2017}%
  \BibitemOpen
  \bibfield  {author} {\bibinfo {author} {\bibfnamefont {P.}~\bibnamefont
  {Siyushev}}, \bibinfo {author} {\bibfnamefont {M.~H.}\ \bibnamefont
  {Metsch}}, \bibinfo {author} {\bibfnamefont {A.}~\bibnamefont {Ijaz}},
  \bibinfo {author} {\bibfnamefont {J.~M.}\ \bibnamefont {Binder}}, \bibinfo
  {author} {\bibfnamefont {M.~K.}\ \bibnamefont {Bhaskar}}, \bibinfo {author}
  {\bibfnamefont {D.~D.}\ \bibnamefont {Sukachev}}, \bibinfo {author}
  {\bibfnamefont {A.}~\bibnamefont {Sipahigil}}, \bibinfo {author}
  {\bibfnamefont {R.~E.}\ \bibnamefont {Evans}}, \bibinfo {author}
  {\bibfnamefont {C.~T.}\ \bibnamefont {Nguyen}}, \bibinfo {author}
  {\bibfnamefont {M.~D.}\ \bibnamefont {Lukin}}, \bibinfo {author}
  {\bibfnamefont {P.~R.}\ \bibnamefont {Hemmer}}, \bibinfo {author}
  {\bibfnamefont {Y.~N.}\ \bibnamefont {Palyanov}}, \bibinfo {author}
  {\bibfnamefont {I.~N.}\ \bibnamefont {Kupriyanov}}, \bibinfo {author}
  {\bibfnamefont {Y.~M.}\ \bibnamefont {Borzdov}}, \bibinfo {author}
  {\bibfnamefont {L.~J.}\ \bibnamefont {Rogers}},\ and\ \bibinfo {author}
  {\bibfnamefont {F.}~\bibnamefont {Jelezko}},\ }\href
  {https://doi.org/10.1103/physrevb.96.081201} {\bibfield  {journal} {\bibinfo
  {journal} {Physical Review B}\ }\textbf {\bibinfo {volume} {96}},\ \bibinfo
  {pages} {081201} (\bibinfo {year} {2017})},\ \Eprint
  {https://arxiv.org/abs/1612.02947} {1612.02947} \BibitemShut {NoStop}%
\bibitem [{\citenamefont {Becker}\ \emph {et~al.}(2018)\citenamefont {Becker},
  \citenamefont {Pingault}, \citenamefont {Groß}, \citenamefont {Gündoğan},
  \citenamefont {Kukharchyk}, \citenamefont {Markham}, \citenamefont {Edmonds},
  \citenamefont {Atatüre}, \citenamefont {Bushev},\ and\ \citenamefont
  {Becher}}]{Becker2018}%
  \BibitemOpen
  \bibfield  {author} {\bibinfo {author} {\bibfnamefont {J.~N.}\ \bibnamefont
  {Becker}}, \bibinfo {author} {\bibfnamefont {B.}~\bibnamefont {Pingault}},
  \bibinfo {author} {\bibfnamefont {D.}~\bibnamefont {Groß}}, \bibinfo
  {author} {\bibfnamefont {M.}~\bibnamefont {Gündoğan}}, \bibinfo {author}
  {\bibfnamefont {N.}~\bibnamefont {Kukharchyk}}, \bibinfo {author}
  {\bibfnamefont {M.}~\bibnamefont {Markham}}, \bibinfo {author} {\bibfnamefont
  {A.}~\bibnamefont {Edmonds}}, \bibinfo {author} {\bibfnamefont
  {M.}~\bibnamefont {Atatüre}}, \bibinfo {author} {\bibfnamefont
  {P.}~\bibnamefont {Bushev}},\ and\ \bibinfo {author} {\bibfnamefont
  {C.}~\bibnamefont {Becher}},\ }\href
  {https://doi.org/10.1103/physrevlett.120.053603} {\bibfield  {journal}
  {\bibinfo  {journal} {Physical Review Letters}\ }\textbf {\bibinfo {volume}
  {120}},\ \bibinfo {pages} {053603} (\bibinfo {year} {2018})},\ \Eprint
  {https://arxiv.org/abs/1708.08263} {1708.08263} \BibitemShut {NoStop}%
\bibitem [{\citenamefont {Weinzetl}\ \emph {et~al.}(2019)\citenamefont
  {Weinzetl}, \citenamefont {Görlitz}, \citenamefont {Becker}, \citenamefont
  {Walmsley}, \citenamefont {Poem}, \citenamefont {Nunn},\ and\ \citenamefont
  {Becher}}]{Weinzetl2019}%
  \BibitemOpen
  \bibfield  {author} {\bibinfo {author} {\bibfnamefont {C.}~\bibnamefont
  {Weinzetl}}, \bibinfo {author} {\bibfnamefont {J.}~\bibnamefont {Görlitz}},
  \bibinfo {author} {\bibfnamefont {J.~N.}\ \bibnamefont {Becker}}, \bibinfo
  {author} {\bibfnamefont {I.~A.}\ \bibnamefont {Walmsley}}, \bibinfo {author}
  {\bibfnamefont {E.}~\bibnamefont {Poem}}, \bibinfo {author} {\bibfnamefont
  {J.}~\bibnamefont {Nunn}},\ and\ \bibinfo {author} {\bibfnamefont
  {C.}~\bibnamefont {Becher}},\ }\href
  {https://doi.org/10.1103/physrevlett.122.063601} {\bibfield  {journal}
  {\bibinfo  {journal} {Physical Review Letters}\ }\textbf {\bibinfo {volume}
  {122}},\ \bibinfo {pages} {063601} (\bibinfo {year} {2019})},\ \Eprint
  {https://arxiv.org/abs/1805.12227} {1805.12227} \BibitemShut {NoStop}%
\bibitem [{\citenamefont {Debroux}\ \emph {et~al.}(2021)\citenamefont
  {Debroux}, \citenamefont {Michaels}, \citenamefont {Purser}, \citenamefont
  {Wan}, \citenamefont {Trusheim}, \citenamefont {Martínez}, \citenamefont
  {Parker}, \citenamefont {Stramma}, \citenamefont {Chen}, \citenamefont
  {Santis}, \citenamefont {Alexeev}, \citenamefont {Ferrari}, \citenamefont
  {Englund}, \citenamefont {Gangloff},\ and\ \citenamefont
  {Atatüre}}]{debroux2021}%
  \BibitemOpen
  \bibfield  {author} {\bibinfo {author} {\bibfnamefont {R.}~\bibnamefont
  {Debroux}}, \bibinfo {author} {\bibfnamefont {C.~P.}\ \bibnamefont
  {Michaels}}, \bibinfo {author} {\bibfnamefont {C.~M.}\ \bibnamefont
  {Purser}}, \bibinfo {author} {\bibfnamefont {N.}~\bibnamefont {Wan}},
  \bibinfo {author} {\bibfnamefont {M.~E.}\ \bibnamefont {Trusheim}}, \bibinfo
  {author} {\bibfnamefont {J.~A.}\ \bibnamefont {Martínez}}, \bibinfo {author}
  {\bibfnamefont {R.~A.}\ \bibnamefont {Parker}}, \bibinfo {author}
  {\bibfnamefont {A.~M.}\ \bibnamefont {Stramma}}, \bibinfo {author}
  {\bibfnamefont {K.~C.}\ \bibnamefont {Chen}}, \bibinfo {author}
  {\bibfnamefont {L.~d.}\ \bibnamefont {Santis}}, \bibinfo {author}
  {\bibfnamefont {E.~M.}\ \bibnamefont {Alexeev}}, \bibinfo {author}
  {\bibfnamefont {A.~C.}\ \bibnamefont {Ferrari}}, \bibinfo {author}
  {\bibfnamefont {D.}~\bibnamefont {Englund}}, \bibinfo {author} {\bibfnamefont
  {D.~A.}\ \bibnamefont {Gangloff}},\ and\ \bibinfo {author} {\bibfnamefont
  {M.}~\bibnamefont {Atatüre}},\ }\href
  {https://doi.org/10.1103/physrevx.11.041041} {\bibfield  {journal} {\bibinfo
  {journal} {Physical Review X}\ }\textbf {\bibinfo {volume} {11}},\ \bibinfo
  {pages} {041041} (\bibinfo {year} {2021})},\ \Eprint
  {https://arxiv.org/abs/2106.00723} {2106.00723} \BibitemShut {NoStop}%
\bibitem [{\citenamefont {Occelli}\ \emph {et~al.}(2003)\citenamefont
  {Occelli}, \citenamefont {Loubeyre},\ and\ \citenamefont
  {Letoullec}}]{Occelli2003}%
  \BibitemOpen
  \bibfield  {author} {\bibinfo {author} {\bibfnamefont {F.}~\bibnamefont
  {Occelli}}, \bibinfo {author} {\bibfnamefont {P.}~\bibnamefont {Loubeyre}},\
  and\ \bibinfo {author} {\bibfnamefont {R.}~\bibnamefont {Letoullec}},\ }\href
  {https://doi.org/10.1038/nmat831} {\bibfield  {journal} {\bibinfo  {journal}
  {Nature Materials}\ }\textbf {\bibinfo {volume} {2}},\ \bibinfo {pages} {151}
  (\bibinfo {year} {2003})}\BibitemShut {NoStop}%
\bibitem [{\citenamefont {Feudis}\ \emph {et~al.}(2020)\citenamefont {Feudis},
  \citenamefont {Tallaire}, \citenamefont {Nicolas}, \citenamefont {Brinza},
  \citenamefont {Goldner}, \citenamefont {H{\'e}tet}, \citenamefont
  {B{\'e}n{\'e}dic},\ and\ \citenamefont {Achard}}]{Feudis2020}%
  \BibitemOpen
  \bibfield  {author} {\bibinfo {author} {\bibfnamefont {M.~D.}\ \bibnamefont
  {Feudis}}, \bibinfo {author} {\bibfnamefont {A.}~\bibnamefont {Tallaire}},
  \bibinfo {author} {\bibfnamefont {L.}~\bibnamefont {Nicolas}}, \bibinfo
  {author} {\bibfnamefont {O.}~\bibnamefont {Brinza}}, \bibinfo {author}
  {\bibfnamefont {P.}~\bibnamefont {Goldner}}, \bibinfo {author} {\bibfnamefont
  {G.}~\bibnamefont {H{\'e}tet}}, \bibinfo {author} {\bibfnamefont
  {F.}~\bibnamefont {B{\'e}n{\'e}dic}},\ and\ \bibinfo {author} {\bibfnamefont
  {J.}~\bibnamefont {Achard}},\ }\href {https://doi.org/10.1002/admi.201901408}
  {\bibfield  {journal} {\bibinfo  {journal} {Advanced Materials Interfaces}\
  }\textbf {\bibinfo {volume} {7}},\ \bibinfo {pages} {1901408} (\bibinfo
  {year} {2020})}\BibitemShut {NoStop}%
\bibitem [{\citenamefont {Tallaire}\ \emph {et~al.}(2019)\citenamefont
  {Tallaire}, \citenamefont {Brinza}, \citenamefont {De~Feudis}, \citenamefont
  {Ferrier}, \citenamefont {Touati}, \citenamefont {Binet}, \citenamefont
  {Nicolas}, \citenamefont {Delord}, \citenamefont {H{\'e}tet}, \citenamefont
  {Herzig}, \citenamefont {Pezzagna}, \citenamefont {Goldner},\ and\
  \citenamefont {Achard}}]{tallaire2019}%
  \BibitemOpen
  \bibfield  {author} {\bibinfo {author} {\bibfnamefont {A.}~\bibnamefont
  {Tallaire}}, \bibinfo {author} {\bibfnamefont {O.}~\bibnamefont {Brinza}},
  \bibinfo {author} {\bibfnamefont {M.}~\bibnamefont {De~Feudis}}, \bibinfo
  {author} {\bibfnamefont {A.}~\bibnamefont {Ferrier}}, \bibinfo {author}
  {\bibfnamefont {N.}~\bibnamefont {Touati}}, \bibinfo {author} {\bibfnamefont
  {L.}~\bibnamefont {Binet}}, \bibinfo {author} {\bibfnamefont
  {L.}~\bibnamefont {Nicolas}}, \bibinfo {author} {\bibfnamefont
  {T.}~\bibnamefont {Delord}}, \bibinfo {author} {\bibfnamefont
  {G.}~\bibnamefont {H{\'e}tet}}, \bibinfo {author} {\bibfnamefont
  {T.}~\bibnamefont {Herzig}}, \bibinfo {author} {\bibfnamefont
  {S.}~\bibnamefont {Pezzagna}}, \bibinfo {author} {\bibfnamefont
  {P.}~\bibnamefont {Goldner}},\ and\ \bibinfo {author} {\bibfnamefont
  {J.}~\bibnamefont {Achard}},\ }\href {https://doi.org/10.1021/acsanm.9b01395}
  {\bibfield  {journal} {\bibinfo  {journal} {ACS Applied Nano Materials}\
  }\textbf {\bibinfo {volume} {2}},\ \bibinfo {pages} {5952} (\bibinfo {year}
  {2019})}\BibitemShut {NoStop}%
\bibitem [{\citenamefont {Boehler}\ and\ \citenamefont
  {Hantsetters}(2004)}]{boehleralmax2004}%
  \BibitemOpen
  \bibfield  {author} {\bibinfo {author} {\bibfnamefont {R.}~\bibnamefont
  {Boehler}}\ and\ \bibinfo {author} {\bibfnamefont {K.~D.}\ \bibnamefont
  {Hantsetters}},\ }\href {https://doi.org/10.1080/08957950412331323924}
  {\bibfield  {journal} {\bibinfo  {journal} {High Pressure Research}\ }\textbf
  {\bibinfo {volume} {24}},\ \bibinfo {pages} {391} (\bibinfo {year}
  {2004})}\BibitemShut {NoStop}%
\bibitem [{\citenamefont {Vos}\ \emph {et~al.}(1991)\citenamefont {Vos},
  \citenamefont {Schouten}, \citenamefont {Young},\ and\ \citenamefont
  {Ross}}]{neonhighpressure1991}%
  \BibitemOpen
  \bibfield  {author} {\bibinfo {author} {\bibfnamefont {W.~L.}\ \bibnamefont
  {Vos}}, \bibinfo {author} {\bibfnamefont {J.~A.}\ \bibnamefont {Schouten}},
  \bibinfo {author} {\bibfnamefont {D.~A.}\ \bibnamefont {Young}},\ and\
  \bibinfo {author} {\bibfnamefont {M.}~\bibnamefont {Ross}},\ }\href
  {https://doi.org/10.1063/1.460683} {\bibfield  {journal} {\bibinfo  {journal}
  {The Journal of Chemical Physics}\ }\textbf {\bibinfo {volume} {94}},\
  \bibinfo {pages} {3835} (\bibinfo {year} {1991})}\BibitemShut {NoStop}%
\bibitem [{\citenamefont {Letoullec}\ \emph {et~al.}(1988)\citenamefont
  {Letoullec}, \citenamefont {Pinceaux},\ and\ \citenamefont
  {Loubeyre}}]{letoullec1988}%
  \BibitemOpen
  \bibfield  {author} {\bibinfo {author} {\bibfnamefont {R.}~\bibnamefont
  {Letoullec}}, \bibinfo {author} {\bibfnamefont {J.~P.}\ \bibnamefont
  {Pinceaux}},\ and\ \bibinfo {author} {\bibfnamefont {P.}~\bibnamefont
  {Loubeyre}},\ }\href {https://doi.org/10.1080/08957958808202482} {\bibfield
  {journal} {\bibinfo  {journal} {High Pressure Research}\ }\textbf {\bibinfo
  {volume} {1}},\ \bibinfo {pages} {77} (\bibinfo {year} {1988})}\BibitemShut
  {NoStop}%
\bibitem [{\citenamefont {Shen}\ \emph {et~al.}(2020)\citenamefont {Shen},
  \citenamefont {Wang}, \citenamefont {Dewaele}, \citenamefont {Wu},
  \citenamefont {Fratanduono}, \citenamefont {Eggert}, \citenamefont {Klotz},
  \citenamefont {Dziubek}, \citenamefont {Loubeyre}, \citenamefont
  {Fat’yanov}, \citenamefont {Asimow}, \citenamefont {Mashimo}, \citenamefont
  {Wentzcovitch},\ and\ \citenamefont {other members of~the
  IPPS~task}}]{shen2020}%
  \BibitemOpen
  \bibfield  {author} {\bibinfo {author} {\bibfnamefont {G.}~\bibnamefont
  {Shen}}, \bibinfo {author} {\bibfnamefont {Y.}~\bibnamefont {Wang}}, \bibinfo
  {author} {\bibfnamefont {A.}~\bibnamefont {Dewaele}}, \bibinfo {author}
  {\bibfnamefont {C.}~\bibnamefont {Wu}}, \bibinfo {author} {\bibfnamefont
  {D.~E.}\ \bibnamefont {Fratanduono}}, \bibinfo {author} {\bibfnamefont
  {J.}~\bibnamefont {Eggert}}, \bibinfo {author} {\bibfnamefont
  {S.}~\bibnamefont {Klotz}}, \bibinfo {author} {\bibfnamefont {K.~F.}\
  \bibnamefont {Dziubek}}, \bibinfo {author} {\bibfnamefont {P.}~\bibnamefont
  {Loubeyre}}, \bibinfo {author} {\bibfnamefont {O.~V.}\ \bibnamefont
  {Fat’yanov}}, \bibinfo {author} {\bibfnamefont {P.~D.}\ \bibnamefont
  {Asimow}}, \bibinfo {author} {\bibfnamefont {T.}~\bibnamefont {Mashimo}},
  \bibinfo {author} {\bibfnamefont {R.~M.~M.}\ \bibnamefont {Wentzcovitch}},\
  and\ \bibinfo {author} {\bibnamefont {other members of~the IPPS~task}},\
  }\href {https://doi.org/10.1080/08957959.2020.1791107} {\bibfield  {journal}
  {\bibinfo  {journal} {High Pressure Research}\ }\textbf {\bibinfo {volume}
  {40}},\ \bibinfo {pages} {1} (\bibinfo {year} {2020})}\BibitemShut {NoStop}%
\bibitem [{\citenamefont {Akahama}\ and\ \citenamefont
  {Kawamura}(2004)}]{akahama2004}%
  \BibitemOpen
  \bibfield  {author} {\bibinfo {author} {\bibfnamefont {Y.}~\bibnamefont
  {Akahama}}\ and\ \bibinfo {author} {\bibfnamefont {H.}~\bibnamefont
  {Kawamura}},\ }\href {https://doi.org/10.1063/1.1778482} {\bibfield
  {journal} {\bibinfo  {journal} {Journal of Applied Physics}\ }\textbf
  {\bibinfo {volume} {96}},\ \bibinfo {pages} {3748} (\bibinfo {year}
  {2004})}\BibitemShut {NoStop}%
\bibitem [{\citenamefont {Lyapin}\ \emph
  {et~al.}(2018{\natexlab{a}})\citenamefont {Lyapin}, \citenamefont {Ilichev},
  \citenamefont {Novikov}, \citenamefont {Davydov},\ and\ \citenamefont
  {Agafonov}}]{lyapin2018}%
  \BibitemOpen
  \bibfield  {author} {\bibinfo {author} {\bibfnamefont {S.}~\bibnamefont
  {Lyapin}}, \bibinfo {author} {\bibfnamefont {I.}~\bibnamefont {Ilichev}},
  \bibinfo {author} {\bibfnamefont {A.}~\bibnamefont {Novikov}}, \bibinfo
  {author} {\bibfnamefont {V.}~\bibnamefont {Davydov}},\ and\ \bibinfo {author}
  {\bibfnamefont {V.}~\bibnamefont {Agafonov}},\ }\href
  {https://doi.org/10.17586/2220-8054-2018-9-1-55-57} {\bibfield  {journal}
  {\bibinfo  {journal} {Nanosystems: Physics, Chemistry, Mathematics}\ }\textbf
  {\bibinfo {volume} {9}},\ \bibinfo {pages} {55} (\bibinfo {year}
  {2018}{\natexlab{a}})}\BibitemShut {NoStop}%
\bibitem [{\citenamefont {Meesala}\ \emph {et~al.}(2018)\citenamefont
  {Meesala}, \citenamefont {Sohn}, \citenamefont {Pingault}, \citenamefont
  {Shao}, \citenamefont {Atikian}, \citenamefont {Holzgrafe}, \citenamefont
  {G{\"u}ndo{\u g}an}, \citenamefont {Stavrakas}, \citenamefont {Sipahigil},
  \citenamefont {Chia}, \citenamefont {Evans}, \citenamefont {Burek},
  \citenamefont {Zhang}, \citenamefont {Wu}, \citenamefont {Pacheco},
  \citenamefont {Abraham}, \citenamefont {Bielejec}, \citenamefont {Lukin},
  \citenamefont {Atat{\"u}re},\ and\ \citenamefont {Lon{\v
  c}ar}}]{meesala2018}%
  \BibitemOpen
  \bibfield  {author} {\bibinfo {author} {\bibfnamefont {S.}~\bibnamefont
  {Meesala}}, \bibinfo {author} {\bibfnamefont {Y.-I.}\ \bibnamefont {Sohn}},
  \bibinfo {author} {\bibfnamefont {B.}~\bibnamefont {Pingault}}, \bibinfo
  {author} {\bibfnamefont {L.}~\bibnamefont {Shao}}, \bibinfo {author}
  {\bibfnamefont {H.~A.}\ \bibnamefont {Atikian}}, \bibinfo {author}
  {\bibfnamefont {J.}~\bibnamefont {Holzgrafe}}, \bibinfo {author}
  {\bibfnamefont {M.}~\bibnamefont {G{\"u}ndo{\u g}an}}, \bibinfo {author}
  {\bibfnamefont {C.}~\bibnamefont {Stavrakas}}, \bibinfo {author}
  {\bibfnamefont {A.}~\bibnamefont {Sipahigil}}, \bibinfo {author}
  {\bibfnamefont {C.}~\bibnamefont {Chia}}, \bibinfo {author} {\bibfnamefont
  {R.}~\bibnamefont {Evans}}, \bibinfo {author} {\bibfnamefont {M.~J.}\
  \bibnamefont {Burek}}, \bibinfo {author} {\bibfnamefont {M.}~\bibnamefont
  {Zhang}}, \bibinfo {author} {\bibfnamefont {L.}~\bibnamefont {Wu}}, \bibinfo
  {author} {\bibfnamefont {J.~L.}\ \bibnamefont {Pacheco}}, \bibinfo {author}
  {\bibfnamefont {J.}~\bibnamefont {Abraham}}, \bibinfo {author} {\bibfnamefont
  {E.}~\bibnamefont {Bielejec}}, \bibinfo {author} {\bibfnamefont {M.~D.}\
  \bibnamefont {Lukin}}, \bibinfo {author} {\bibfnamefont {M.}~\bibnamefont
  {Atat{\"u}re}},\ and\ \bibinfo {author} {\bibfnamefont {M.}~\bibnamefont
  {Lon{\v c}ar}},\ }\href {https://doi.org/10.1103/PhysRevB.97.205444}
  {\bibfield  {journal} {\bibinfo  {journal} {Physical Review B}\ }\textbf
  {\bibinfo {volume} {97}},\ \bibinfo {pages} {205444} (\bibinfo {year}
  {2018})}\BibitemShut {NoStop}%
\bibitem [{\citenamefont {Lindner}\ \emph {et~al.}(2018)\citenamefont
  {Lindner}, \citenamefont {Bommer}, \citenamefont {Muzha}, \citenamefont
  {Krueger}, \citenamefont {Gines}, \citenamefont {Mandal}, \citenamefont
  {Williams}, \citenamefont {Londero}, \citenamefont {Gali},\ and\
  \citenamefont {Becher}}]{lindner2018}%
  \BibitemOpen
  \bibfield  {author} {\bibinfo {author} {\bibfnamefont {S.}~\bibnamefont
  {Lindner}}, \bibinfo {author} {\bibfnamefont {A.}~\bibnamefont {Bommer}},
  \bibinfo {author} {\bibfnamefont {A.}~\bibnamefont {Muzha}}, \bibinfo
  {author} {\bibfnamefont {A.}~\bibnamefont {Krueger}}, \bibinfo {author}
  {\bibfnamefont {L.}~\bibnamefont {Gines}}, \bibinfo {author} {\bibfnamefont
  {S.}~\bibnamefont {Mandal}}, \bibinfo {author} {\bibfnamefont
  {O.}~\bibnamefont {Williams}}, \bibinfo {author} {\bibfnamefont
  {E.}~\bibnamefont {Londero}}, \bibinfo {author} {\bibfnamefont
  {A.}~\bibnamefont {Gali}},\ and\ \bibinfo {author} {\bibfnamefont
  {C.}~\bibnamefont {Becher}},\ }\href
  {https://doi.org/10.1088/1367-2630/aae93f} {\bibfield  {journal} {\bibinfo
  {journal} {New Journal of Physics}\ }\textbf {\bibinfo {volume} {20}},\
  \bibinfo {pages} {115002} (\bibinfo {year} {2018})}\BibitemShut {NoStop}%
\bibitem [{\citenamefont {Davydov}\ \emph {et~al.}(2014)\citenamefont
  {Davydov}, \citenamefont {Rakhmanina}, \citenamefont {Lyapin}, \citenamefont
  {Ilichev}, \citenamefont {Boldyrev}, \citenamefont {Shiryaev},\ and\
  \citenamefont {Agafonov}}]{davydov2014}%
  \BibitemOpen
  \bibfield  {author} {\bibinfo {author} {\bibfnamefont {V.~A.}\ \bibnamefont
  {Davydov}}, \bibinfo {author} {\bibfnamefont {A.~V.}\ \bibnamefont
  {Rakhmanina}}, \bibinfo {author} {\bibfnamefont {S.~G.}\ \bibnamefont
  {Lyapin}}, \bibinfo {author} {\bibfnamefont {I.~D.}\ \bibnamefont {Ilichev}},
  \bibinfo {author} {\bibfnamefont {K.~N.}\ \bibnamefont {Boldyrev}}, \bibinfo
  {author} {\bibfnamefont {A.~A.}\ \bibnamefont {Shiryaev}},\ and\ \bibinfo
  {author} {\bibfnamefont {V.~N.}\ \bibnamefont {Agafonov}},\ }\href
  {https://doi.org/10.1134/S002136401410004X} {\bibfield  {journal} {\bibinfo
  {journal} {JETP Letters}\ }\textbf {\bibinfo {volume} {99}},\ \bibinfo
  {pages} {585} (\bibinfo {year} {2014})}\BibitemShut {NoStop}%
\bibitem [{\citenamefont {Nguyen}\ \emph {et~al.}(2018)\citenamefont {Nguyen},
  \citenamefont {Evans}, \citenamefont {Sipahigil}, \citenamefont {Bhaskar},
  \citenamefont {Sukachev}, \citenamefont {Agafonov}, \citenamefont {Davydov},
  \citenamefont {Kulikova}, \citenamefont {Jelezko},\ and\ \citenamefont
  {Lukin}}]{chen2018}%
  \BibitemOpen
  \bibfield  {author} {\bibinfo {author} {\bibfnamefont {C.~T.}\ \bibnamefont
  {Nguyen}}, \bibinfo {author} {\bibfnamefont {R.~E.}\ \bibnamefont {Evans}},
  \bibinfo {author} {\bibfnamefont {A.}~\bibnamefont {Sipahigil}}, \bibinfo
  {author} {\bibfnamefont {M.~K.}\ \bibnamefont {Bhaskar}}, \bibinfo {author}
  {\bibfnamefont {D.~D.}\ \bibnamefont {Sukachev}}, \bibinfo {author}
  {\bibfnamefont {V.~N.}\ \bibnamefont {Agafonov}}, \bibinfo {author}
  {\bibfnamefont {V.~A.}\ \bibnamefont {Davydov}}, \bibinfo {author}
  {\bibfnamefont {L.~F.}\ \bibnamefont {Kulikova}}, \bibinfo {author}
  {\bibfnamefont {F.}~\bibnamefont {Jelezko}},\ and\ \bibinfo {author}
  {\bibfnamefont {M.~D.}\ \bibnamefont {Lukin}},\ }\href
  {https://doi.org/10.1063/1.5029904} {\bibfield  {journal} {\bibinfo
  {journal} {Applied Physics Letters}\ }\textbf {\bibinfo {volume} {112}},\
  \bibinfo {pages} {203102} (\bibinfo {year} {2018})},\ \Eprint
  {https://arxiv.org/abs/1708.05419} {1708.05419} \BibitemShut {NoStop}%
\bibitem [{\citenamefont {Isa}\ \emph {et~al.}(2022)\citenamefont {Isa},
  \citenamefont {Joliffe}, \citenamefont {Wouterlood}, \citenamefont {Ho},
  \citenamefont {Volz}, \citenamefont {Bendavid},\ and\ \citenamefont
  {Rogers}}]{Isa2022}%
  \BibitemOpen
  \bibfield  {author} {\bibinfo {author} {\bibfnamefont {F.}~\bibnamefont
  {Isa}}, \bibinfo {author} {\bibfnamefont {M.}~\bibnamefont {Joliffe}},
  \bibinfo {author} {\bibfnamefont {B.}~\bibnamefont {Wouterlood}}, \bibinfo
  {author} {\bibfnamefont {N.~H.}\ \bibnamefont {Ho}}, \bibinfo {author}
  {\bibfnamefont {T.}~\bibnamefont {Volz}}, \bibinfo {author} {\bibfnamefont
  {A.}~\bibnamefont {Bendavid}},\ and\ \bibinfo {author} {\bibfnamefont
  {L.~J.}\ \bibnamefont {Rogers}},\ }\href@noop {} {\bibfield  {journal}
  {\bibinfo  {journal} {arXiv}\ } (\bibinfo {year} {2022})},\ \Eprint
  {https://arxiv.org/abs/2208.08075} {2208.08075} \BibitemShut {NoStop}%
\bibitem [{\citenamefont {Hughes}\ and\ \citenamefont
  {Runciman}(1967)}]{hughes1967}%
  \BibitemOpen
  \bibfield  {author} {\bibinfo {author} {\bibfnamefont {A.~E.}\ \bibnamefont
  {Hughes}}\ and\ \bibinfo {author} {\bibfnamefont {W.~A.}\ \bibnamefont
  {Runciman}},\ }\href {https://doi.org/10.1088/0370-1328/90/3/328} {\bibfield
  {journal} {\bibinfo  {journal} {Proceedings of the Physical Society}\
  }\textbf {\bibinfo {volume} {90}},\ \bibinfo {pages} {827} (\bibinfo {year}
  {1967})}\BibitemShut {NoStop}%
\bibitem [{\citenamefont {Hepp}\ \emph {et~al.}(2014)\citenamefont {Hepp},
  \citenamefont {M{\"u}ller}, \citenamefont {Waselowski}, \citenamefont
  {Becker}, \citenamefont {Pingault}, \citenamefont {Sternschulte},
  \citenamefont {{Steinm{\"u}ller-Nethl}}, \citenamefont {Gali}, \citenamefont
  {Maze}, \citenamefont {Atat{\"u}re},\ and\ \citenamefont
  {Becher}}]{hepp2014}%
  \BibitemOpen
  \bibfield  {author} {\bibinfo {author} {\bibfnamefont {C.}~\bibnamefont
  {Hepp}}, \bibinfo {author} {\bibfnamefont {T.}~\bibnamefont {M{\"u}ller}},
  \bibinfo {author} {\bibfnamefont {V.}~\bibnamefont {Waselowski}}, \bibinfo
  {author} {\bibfnamefont {J.~N.}\ \bibnamefont {Becker}}, \bibinfo {author}
  {\bibfnamefont {B.}~\bibnamefont {Pingault}}, \bibinfo {author}
  {\bibfnamefont {H.}~\bibnamefont {Sternschulte}}, \bibinfo {author}
  {\bibfnamefont {D.}~\bibnamefont {{Steinm{\"u}ller-Nethl}}}, \bibinfo
  {author} {\bibfnamefont {A.}~\bibnamefont {Gali}}, \bibinfo {author}
  {\bibfnamefont {J.~R.}\ \bibnamefont {Maze}}, \bibinfo {author}
  {\bibfnamefont {M.}~\bibnamefont {Atat{\"u}re}},\ and\ \bibinfo {author}
  {\bibfnamefont {C.}~\bibnamefont {Becher}},\ }\href
  {https://doi.org/10.1103/PhysRevLett.112.036405} {\bibfield  {journal}
  {\bibinfo  {journal} {Physical Review Letters}\ }\textbf {\bibinfo {volume}
  {112}},\ \bibinfo {pages} {036405} (\bibinfo {year} {2014})}\BibitemShut
  {NoStop}%
\bibitem [{\citenamefont {Grazioso}\ \emph {et~al.}(2013)\citenamefont
  {Grazioso}, \citenamefont {Patton}, \citenamefont {Delaney}, \citenamefont
  {Markham}, \citenamefont {Twitchen},\ and\ \citenamefont
  {Smith}}]{grazioso2013}%
  \BibitemOpen
  \bibfield  {author} {\bibinfo {author} {\bibfnamefont {F.}~\bibnamefont
  {Grazioso}}, \bibinfo {author} {\bibfnamefont {B.~R.}\ \bibnamefont
  {Patton}}, \bibinfo {author} {\bibfnamefont {P.}~\bibnamefont {Delaney}},
  \bibinfo {author} {\bibfnamefont {M.~L.}\ \bibnamefont {Markham}}, \bibinfo
  {author} {\bibfnamefont {D.~J.}\ \bibnamefont {Twitchen}},\ and\ \bibinfo
  {author} {\bibfnamefont {J.~M.}\ \bibnamefont {Smith}},\ }\href
  {https://doi.org/10.1063/1.4819834} {\bibfield  {journal} {\bibinfo
  {journal} {Applied Physics Letters}\ }\textbf {\bibinfo {volume} {103}},\
  \bibinfo {pages} {101905} (\bibinfo {year} {2013})}\BibitemShut {NoStop}%
\bibitem [{\citenamefont {Londero}\ \emph {et~al.}(2018)\citenamefont
  {Londero}, \citenamefont {Thiering}, \citenamefont {Razinkovas},
  \citenamefont {Gali},\ and\ \citenamefont {Alkauskas}}]{londero2018}%
  \BibitemOpen
  \bibfield  {author} {\bibinfo {author} {\bibfnamefont {E.}~\bibnamefont
  {Londero}}, \bibinfo {author} {\bibfnamefont {G.}~\bibnamefont {Thiering}},
  \bibinfo {author} {\bibfnamefont {L.}~\bibnamefont {Razinkovas}}, \bibinfo
  {author} {\bibfnamefont {A.}~\bibnamefont {Gali}},\ and\ \bibinfo {author}
  {\bibfnamefont {A.}~\bibnamefont {Alkauskas}},\ }\href
  {https://doi.org/10.1103/PhysRevB.98.035306} {\bibfield  {journal} {\bibinfo
  {journal} {Phys. Rev. B}\ }\textbf {\bibinfo {volume} {98}},\ \bibinfo
  {pages} {035306} (\bibinfo {year} {2018})}\BibitemShut {NoStop}%
\bibitem [{\citenamefont {Sun}\ \emph {et~al.}(2015)\citenamefont {Sun},
  \citenamefont {Ruzsinszky},\ and\ \citenamefont {Perdew}}]{scan}%
  \BibitemOpen
  \bibfield  {author} {\bibinfo {author} {\bibfnamefont {J.}~\bibnamefont
  {Sun}}, \bibinfo {author} {\bibfnamefont {A.}~\bibnamefont {Ruzsinszky}},\
  and\ \bibinfo {author} {\bibfnamefont {J.~P.}\ \bibnamefont {Perdew}},\
  }\href {https://doi.org/10.1103/PhysRevLett.115.036402} {\bibfield  {journal}
  {\bibinfo  {journal} {Phys. Rev. Lett.}\ }\textbf {\bibinfo {volume} {115}},\
  \bibinfo {pages} {036402} (\bibinfo {year} {2015})}\BibitemShut {NoStop}%
\bibitem [{\citenamefont {Vinet}\ \emph {et~al.}(1987)\citenamefont {Vinet},
  \citenamefont {Smith}, \citenamefont {Ferrante},\ and\ \citenamefont
  {Rose}}]{eos}%
  \BibitemOpen
  \bibfield  {author} {\bibinfo {author} {\bibfnamefont {P.}~\bibnamefont
  {Vinet}}, \bibinfo {author} {\bibfnamefont {J.~R.}\ \bibnamefont {Smith}},
  \bibinfo {author} {\bibfnamefont {J.}~\bibnamefont {Ferrante}},\ and\
  \bibinfo {author} {\bibfnamefont {J.~H.}\ \bibnamefont {Rose}},\ }\href
  {https://doi.org/10.1103/PhysRevB.35.1945} {\bibfield  {journal} {\bibinfo
  {journal} {Phys. Rev. B}\ }\textbf {\bibinfo {volume} {35}},\ \bibinfo
  {pages} {1945} (\bibinfo {year} {1987})}\BibitemShut {NoStop}%
\bibitem [{\citenamefont {Hao}\ \emph {et~al.}(2012)\citenamefont {Hao},
  \citenamefont {Fang}, \citenamefont {Sun}, \citenamefont {Csonka},
  \citenamefont {Philipsen},\ and\ \citenamefont {Perdew}}]{pan2012a}%
  \BibitemOpen
  \bibfield  {author} {\bibinfo {author} {\bibfnamefont {P.}~\bibnamefont
  {Hao}}, \bibinfo {author} {\bibfnamefont {Y.}~\bibnamefont {Fang}}, \bibinfo
  {author} {\bibfnamefont {J.}~\bibnamefont {Sun}}, \bibinfo {author}
  {\bibfnamefont {G.~I.}\ \bibnamefont {Csonka}}, \bibinfo {author}
  {\bibfnamefont {P.~H.~T.}\ \bibnamefont {Philipsen}},\ and\ \bibinfo {author}
  {\bibfnamefont {J.~P.}\ \bibnamefont {Perdew}},\ }\href
  {https://doi.org/10.1103/PhysRevB.85.014111} {\bibfield  {journal} {\bibinfo
  {journal} {Physical Review B}\ }\textbf {\bibinfo {volume} {85}},\ \bibinfo
  {pages} {014111} (\bibinfo {year} {2012})}\BibitemShut {NoStop}%
\bibitem [{\citenamefont {Zouboulis}\ \emph {et~al.}(1998)\citenamefont
  {Zouboulis}, \citenamefont {Grimsditch}, \citenamefont {Ramdas},\ and\
  \citenamefont {Rodriguez}}]{zouboulis1998}%
  \BibitemOpen
  \bibfield  {author} {\bibinfo {author} {\bibfnamefont {E.~S.}\ \bibnamefont
  {Zouboulis}}, \bibinfo {author} {\bibfnamefont {M.}~\bibnamefont
  {Grimsditch}}, \bibinfo {author} {\bibfnamefont {A.~K.}\ \bibnamefont
  {Ramdas}},\ and\ \bibinfo {author} {\bibfnamefont {S.}~\bibnamefont
  {Rodriguez}},\ }\href {https://doi.org/10.1103/PhysRevB.57.2889} {\bibfield
  {journal} {\bibinfo  {journal} {Physical Review B}\ }\textbf {\bibinfo
  {volume} {57}},\ \bibinfo {pages} {2889} (\bibinfo {year}
  {1998})}\BibitemShut {NoStop}%
\bibitem [{\citenamefont {Bl{\"o}chl}(1994)}]{paw}%
  \BibitemOpen
  \bibfield  {author} {\bibinfo {author} {\bibfnamefont {P.~E.}\ \bibnamefont
  {Bl{\"o}chl}},\ }\href@noop {} {\bibfield  {journal} {\bibinfo  {journal}
  {Physical review B}\ }\textbf {\bibinfo {volume} {50}},\ \bibinfo {pages}
  {17953} (\bibinfo {year} {1994})}\BibitemShut {NoStop}%
\bibitem [{\citenamefont {Kresse}\ and\ \citenamefont
  {Furthm\"uller}(1996)}]{vasp}%
  \BibitemOpen
  \bibfield  {author} {\bibinfo {author} {\bibfnamefont {G.}~\bibnamefont
  {Kresse}}\ and\ \bibinfo {author} {\bibfnamefont {J.}~\bibnamefont
  {Furthm\"uller}},\ }\href {https://doi.org/10.1103/PhysRevB.54.11169}
  {\bibfield  {journal} {\bibinfo  {journal} {Phys. Rev. B}\ }\textbf {\bibinfo
  {volume} {54}},\ \bibinfo {pages} {11169} (\bibinfo {year}
  {1996})}\BibitemShut {NoStop}%
\bibitem [{\citenamefont {Harrison}(2012)}]{harrison2012}%
  \BibitemOpen
  \bibfield  {author} {\bibinfo {author} {\bibfnamefont {W.~A.}\ \bibnamefont
  {Harrison}},\ }\href@noop {} {\emph {\bibinfo {title} {Electronic structure
  and the properties of solids: The physics of the chemical bond}}}\ (\bibinfo
  {publisher} {Courier Corporation},\ \bibinfo {year} {2012})\BibitemShut
  {NoStop}%
\bibitem [{\citenamefont {Rose}\ \emph {et~al.}(2018)\citenamefont {Rose},
  \citenamefont {Huang}, \citenamefont {Zhang}, \citenamefont {Stevenson},
  \citenamefont {Tyryshkin}, \citenamefont {Sangtawesin}, \citenamefont
  {Srinivasan}, \citenamefont {Loudin}, \citenamefont {Markham}, \citenamefont
  {Edmonds}, \citenamefont {Twitchen}, \citenamefont {Lyon},\ and\
  \citenamefont {de~Leon}}]{Rose2018}%
  \BibitemOpen
  \bibfield  {author} {\bibinfo {author} {\bibfnamefont {B.~C.}\ \bibnamefont
  {Rose}}, \bibinfo {author} {\bibfnamefont {D.}~\bibnamefont {Huang}},
  \bibinfo {author} {\bibfnamefont {Z.-H.}\ \bibnamefont {Zhang}}, \bibinfo
  {author} {\bibfnamefont {P.}~\bibnamefont {Stevenson}}, \bibinfo {author}
  {\bibfnamefont {A.~M.}\ \bibnamefont {Tyryshkin}}, \bibinfo {author}
  {\bibfnamefont {S.}~\bibnamefont {Sangtawesin}}, \bibinfo {author}
  {\bibfnamefont {S.}~\bibnamefont {Srinivasan}}, \bibinfo {author}
  {\bibfnamefont {L.}~\bibnamefont {Loudin}}, \bibinfo {author} {\bibfnamefont
  {M.~L.}\ \bibnamefont {Markham}}, \bibinfo {author} {\bibfnamefont {A.~M.}\
  \bibnamefont {Edmonds}}, \bibinfo {author} {\bibfnamefont {D.~J.}\
  \bibnamefont {Twitchen}}, \bibinfo {author} {\bibfnamefont {S.~A.}\
  \bibnamefont {Lyon}},\ and\ \bibinfo {author} {\bibfnamefont {N.~P.}\
  \bibnamefont {de~Leon}},\ }\href {https://doi.org/10.1126/science.aao0290}
  {\bibfield  {journal} {\bibinfo  {journal} {Science}\ }\textbf {\bibinfo
  {volume} {361}},\ \bibinfo {pages} {60} (\bibinfo {year} {2018})},\ \Eprint
  {https://arxiv.org/abs/1706.01555} {1706.01555} \BibitemShut {NoStop}%
\bibitem [{\citenamefont {Vindolet~et al.}(2022)}]{zenodog4v}%
  \BibitemOpen
  \bibfield  {author} {\bibinfo {author} {\bibfnamefont {B.}~\bibnamefont
  {Vindolet~et al.}},\ }\href@noop {} {\bibinfo {title} {{Replication data for:
  Optical properties of SiV and GeV color centers in nanodiamonds under
  hydrostatic pressures up to 180 GPa}}},\ \bibinfo {howpublished} {Zenodo
  \url{https://doi.org/10.5281/zenodo.7086416}} (\bibinfo {year} {2022}),\
  \bibinfo {note} {online; accessed 16 September 2022}\BibitemShut {NoStop}%
\bibitem [{\citenamefont {Lyapin}\ \emph
  {et~al.}(2018{\natexlab{b}})\citenamefont {Lyapin}, \citenamefont {Razgulov},
  \citenamefont {Novikov}, \citenamefont {Ekimov},\ and\ \citenamefont
  {Kondrin}}]{lyapin2018:GeV}%
  \BibitemOpen
  \bibfield  {author} {\bibinfo {author} {\bibfnamefont {S.~G.}\ \bibnamefont
  {Lyapin}}, \bibinfo {author} {\bibfnamefont {A.~A.}\ \bibnamefont
  {Razgulov}}, \bibinfo {author} {\bibfnamefont {A.~P.}\ \bibnamefont
  {Novikov}}, \bibinfo {author} {\bibfnamefont {E.~A.}\ \bibnamefont
  {Ekimov}},\ and\ \bibinfo {author} {\bibfnamefont {M.~V.}\ \bibnamefont
  {Kondrin}},\ }\href {https://doi.org/10.17586/2220-8054-2018-9-1-67-69}
  {\bibfield  {journal} {\bibinfo  {journal} {Nanosystems: Physics, Chemistry,
  Mathematics}\ }\textbf {\bibinfo {volume} {9}},\ \bibinfo {pages} {67}
  (\bibinfo {year} {2018}{\natexlab{b}})}\BibitemShut {NoStop}%
\bibitem [{\citenamefont {Razgulov}\ \emph {et~al.}(2021)\citenamefont
  {Razgulov}, \citenamefont {Lyapin}, \citenamefont {Novikov},\ and\
  \citenamefont {Ekimov}}]{razgulov2021}%
  \BibitemOpen
  \bibfield  {author} {\bibinfo {author} {\bibfnamefont {A.}~\bibnamefont
  {Razgulov}}, \bibinfo {author} {\bibfnamefont {S.}~\bibnamefont {Lyapin}},
  \bibinfo {author} {\bibfnamefont {A.}~\bibnamefont {Novikov}},\ and\ \bibinfo
  {author} {\bibfnamefont {E.}~\bibnamefont {Ekimov}},\ }\href
  {https://doi.org/10.1016/j.diamond.2021.108379} {\bibfield  {journal}
  {\bibinfo  {journal} {Diamond and Related Materials}\ }\textbf {\bibinfo
  {volume} {116}},\ \bibinfo {pages} {108379} (\bibinfo {year}
  {2021})}\BibitemShut {NoStop}%
\bibitem [{\citenamefont {Ekimov}\ \emph {et~al.}(2019)\citenamefont {Ekimov},
  \citenamefont {Lyapin}, \citenamefont {Razgulov},\ and\ \citenamefont
  {Kondrin}}]{ekimov2019}%
  \BibitemOpen
  \bibfield  {author} {\bibinfo {author} {\bibfnamefont {E.~A.}\ \bibnamefont
  {Ekimov}}, \bibinfo {author} {\bibfnamefont {S.~G.}\ \bibnamefont {Lyapin}},
  \bibinfo {author} {\bibfnamefont {A.~A.}\ \bibnamefont {Razgulov}},\ and\
  \bibinfo {author} {\bibfnamefont {M.~V.}\ \bibnamefont {Kondrin}},\ }\href
  {https://doi.org/10.1134/s1063776119090097} {\bibfield  {journal} {\bibinfo
  {journal} {Journal of Experimental and Theoretical Physics}\ }\textbf
  {\bibinfo {volume} {129}},\ \bibinfo {pages} {855} (\bibinfo {year}
  {2019})}\BibitemShut {NoStop}%
\end{thebibliography}%

\end{document}